\documentclass[fleqn,usenatbib]{mnras}

\usepackage[T1]{fontenc}
\usepackage{ae,aecompl}

\usepackage{here}
\usepackage{float}
\usepackage{array, makecell}

\usepackage{graphicx}	
\usepackage{amsmath}	
\usepackage{amssymb}	

\usepackage{newtxtext,newtxmath}

\begin{document}
\label{firstpage}
\pagerange{\pageref{firstpage}--\pageref{lastpage}}

\title[Modelling H$_2$ and its effects on galaxy evolution]{Modelling H$_2$ and its effects on star formation using a joint implementation of {\LARGE GADGET-3} and {\LARGE KROME}}

\author[Sillero et al. 2020]{Emanuel Sillero,$^{1}$\thanks{E-mail: esillero@oac.unc.edu.ar}
Patricia B. Tissera,$^{2,3,4}$
Diego G. Lambas,$^{1}$
Stefano Bovino,$^{5}$
\newauthor Dominik R. Schleicher,$^{5}$
Tommaso Grassi,$^{6,7}$
Gustavo Bruzual,$^{8}$
\& Stéphane Charlot$^{9}$
\\ \\
$^{1}$ Instituto de Astronom\'ia Te\'orica y Experimental (CONICET-UNC), Laprida 925, C\'ordoba, Argentina.\\
$^{2}$ Departamento de Ciencias F\'isicas, Universidad Andres Bello, Fernandez Concha 700, Santiago, Chile. \\
$^{3}$ Correspondient researcher, Instituto de Astronom\'ia Te\'orica y Experimental (CONICET-UNC), Laprida 925, C\'ordoba, Argentina.\\
$^{4}$ Millennium Institute of Astrophysics, Universidad Andres Bello, Fernandez Concha 700, Santiago, Chile.\\
$^{5}$ Departamento de Astronom\'ia, Facultad Ciencias F\'isicas y Matem\'aticas, Universidad de Concepci\'on, Av. Esteban Iturra s/n Barrio Universitario,\\ \ \ \ Casilla 160-C, Concepci\'on, Chile\\
$^{6}$ Universit\"ats-Sternwarte M\"unchen, Scheinerstr. 1, D-81679 M\"unchen, Germany\\
$^{7}$ Excellence Cluster Origin and Structure of the Universe, Boltzmannstr.2, D-85748 Garching bei M\"unchen, Germany\\
$^{8}$ Instituto de Radioastronom\'ia y Astrof\'isica, UNAM, Campus Morelia, Michoac\'an, C.P. 58089, M\'exico\\
$^{9}$ Sorbonne Universit\'e, CNRS, UMR7095, Institut d'Astrophysique de Paris, F-75014, Paris, France
}
\date{Accepted XXX. Received YYY; in original form ZZZ}

\pubyear{2020}

\maketitle

\begin{abstract}
We present {\small P-GADGET3-K}, an updated version of {\small GADGET3}, that incorporates the chemistry package {\small KROME}. {\small P-GADGET3-K} follows the hydrodynamical and chemical evolution of cosmic structures, incorporating the chemistry and cooling of H$_2$ and metal cooling in non-equilibrium. 
We performed different runs of the same ICs to assess the impact of various physical parameters and prescriptions, namely gas metallicity, molecular hydrogen formation on dust, star formation recipes including or not H$_2$ dependence, and the effects of numerical resolution. We find that the characteristics of the simulated systems, both globally and at kpc-scales, are in good agreement with several observable properties of molecular gas in star-forming galaxies.
The surface density profiles of SFR and H$_2$ are found to vary with the clumping factor and resolution. In agreement with previous results, the chemical enrichment of the gas component is found to be a key ingredient to model the formation and distribution of H$_2$ as a function of gas density and temperature. A SF algorithm that takes into account the H$_2$ fraction together with a treatment for  the local stellar radiation field improves the agreement with observed H$_2$ abundances over a wide range of gas densities and with the molecular Kennicutt-Schmidt law, implying a  more realistic modelling of the star formation process.
\end{abstract}



\section{Introduction}

The presence of molecular hydrogen (H$_2$), the most common molecule in the Universe \citep{Herbst2001}, extends to a wide range of scales in galaxies and is essential for cooling the interstellar medium (ISM) \citep{Glover.Abel2008}. At pc-scales, the higher density regions protect molecules from being destroyed, enabling the formation of giant molecular clouds (GMCs). In the Milky Way, the GMC mass distribution follows a power law similar to that of the luminosity distribution of OB stars \citep{Williams.McKee1997}, while their velocity dispersion obeys a power law that increases with the GMCs size \citep{Larson1981}. Meanwhile, observations of nearby galaxies on kpc-scales show that the H$_2$ surface density correlates with the star formation rate (SFR) \citep{Bigiel2008, Schruba2011, Leroy2013}.

\citet{Schmidt1959}, from an analysis in the solar neighbourhood, and \citet{Kennicutt1998}, from a study to external galaxies, demonstrate that the neutral hydrogen surface density correlates with the SFR surface density (KS law). However, new measurements, which resolved the structure of external galaxies and the H$_2$ within them \citep[e.g.][]{Bigiel2008}, find a tighter correlation between H$_2$ and SFR (the so-called molecular KS law). Although \citet{Leroy2013} confirm this 
trend and observations agree that cold and dense GMCs are associated with young and hot stars and active star formation \citep[e.g.][]{Genzel.Stutzki1989}, recent surveys of more than a thousand galaxies report a weak correlation between the total H$_2$ content and the SFR \citep{Saintonge2017, Catinella2018}.
Hence, it is not yet clear if there is causal connection behind the correlation between H$_2$ content and star formation activity (SF). Unfortunately, H$_2$ is challenging to detect directly through observations owing to its lack of a dipole moment. 
This leads to CO measures to infer the amount of H$_2$ and identify the GMCs \citep{Solomon1987}. The H$_2$ abundance can then be estimated by applying a conversion factor, $X_{\mathrm{CO}}$, between the CO intensity and the H$_2$ column density. This conversion factor has been extensively measured and is found to be approximately constant for the Milky Way, $X_{\mathrm{CO}} \approx 2\times10^{20} \mathrm{cm^{-2} K^{-1} km^{-1} s}$ \citep{Solomon1987, Strong.Mattox1996, Dame2001, Goldsmith2008, Pineda2010}. However, studies beyond the Milky Way show that it might depend on galaxy morphology and metallicity \citep{Bolatto2013}. The determination of the conversion factor is key to estimate the H$_2$ content as a function of time.

There are two channels for the formation of H$_2$ in the ISM: through exothermic synthesis reactions in the gas phase or catalysis by dust surface. The former are slow processes which were only important in the early Universe when heavy chemical elements were scarce \citep{Galli.Palla1998}. The later is the more probable formation channel and is based on the use of interstellar dust grains as catalysts by which HI sticks and coalesces into H$_2$ in a more efficient way \citep{Wakelam2017}. Collisions at high temperatures \citep[$T {\sim} 1000$ K][]{Glover.Abel2008}
and ionisation by high-energy photons \citep[$h\nu \ge 15.42$ eV;][]{Abel1997} can dissociate H$_2$. Additionally, photons with energy that falls into the Lyman-Werner (LW) band (11.2 to 13.6 eV) also dissociate H$_2$. 
The GMCs can absorb the LW photons at stronger wavelengths forming a HI protective layer, while weaker wavelengths can penetrate towards the inner regions until they are also absorbed.

Two processes shield H$_2$ from radiation, the shielding by dust and the H$_2$ self-shielding. The absorption rate is highly dependent on the wavelength of the LW band and only about 10 per cent of the LW absorption leads to H$_2$ dissociation \citep{Stecher.Williams1967}. The rest of the absorbed photons do not contribute to photodissociation \citep{Sternberg2014}. On the other hand, the H$_2$ self-shielding is weak at low-column densities and increases further into the cloud.

The era of increasing precision for H$_2$ measurement requires a similar increase in sophistication for modelling the physics of the interstellar medium and H$_2$ chemistry in cosmological simulations. The past decade has seen a number of methods to model the H$_2$ chemistry in both semi-analytical galaxy models and hydrodynamical simulations of galaxy formation. Semi-analytical models use equilibrium equations to find the H$_2$ fraction under the assumption that the chemistry in each volume element is in an equilibrium state determined by purely local variables (e.g. \citet{McKee.Krumholz2010} - KMT model).
On the other hand, hydrodynamic simulations use either equilibrium equations \citep[e.g.][]{Robertson.Kravtsov2008,Kuhlen2012,Hopkins2014} or a series of non-equilibrium chemical networks \citep[e.g.][]{Richings.Schaye2016a,Lupi2018}. Equilibrium calculations have the advantage of being faster using the assumption that the chemical species are in equilibrium with their environment. Non-equilibrium codes instead, use local rates of destruction and creation of chemical species, and networks of rate equations to track locally atomic and molecular hydrogen in their galaxy simulations, together with the thermal evolution.
\citet{Krumholz.Gnedin2011} compared both kind of models, showing that they diverge at low metallicities. Non-equilibrium models are capable of preserving H$_2$ at lower densities \citep{Tomassetti2014}, producing clumpier H$_2$ mass distributions and star forming regions closer to the KS law \citep{Pallottini2017}.
Several works have been reported based on non-equilibrium schemes. \citet{Richings.Schaye2016b} adapt the network developed by \citet{Richings2014a,Richings2014b} for a version of {\small GADGET-3} \citep{Springel2005b}. These authors use a fixed metallicity through out the entire simulation as in \citet{Nickerson2019}. 
\citet{Hu2016,Hu2017} adapt the \citet{Glover.Clark2012} network for {\small GADGET-3} and also model metal production by SNae. They track six species (H$_2$, HII, CO, HI, CII, and OI), of which only the first three are computed explicitly using the chemical network.
Other authors resorted to the chemistry package {\small KROME} \citep{Grassi2014}, which can solve the chemical rate equations for any given reactions network, and also includes many other processes that are tightly connected to astrochemistry. Particularly, {\small KROME} allows to add thermal processes (as cooling from endothermic reactions and from several atoms and molecules, and heating from exothermic chemical reactions and photochemistry) considering, among others factors, the metal enrichment and a radiation field. Several works have implemented this package in hydrodynamical cosmological codes as {\small RAMSES} \citep{Pallottini2017}, {\small GASOLINE2} \citep{Capelo2018} and {\small GIZMO} \citep{Lupi2018}.
These works assume different  spatial uniform initial metallicity for the simulated galaxy.

Overall, the adopted chemical networks for the formation and destruction of H$_2$ are similar in all the mentioned works. They differ principally in how they model the synthesis of H$_2$ by dust grains. The first option is whether to include a clumping factor to account for unresolved dense structures in molecular clouds. The clumping factor enhances the formation of H$_2$ by some factor. The most common option is to adopt a constant factor \citep{Christensen2012,Capelo2018}. However, \citet{Micic2012} find that this may lead to an overproduction of H$_2$ in high-density regions. Instead, several works explore variable clumping factors. For example, \citet{Tomassetti2014} report that their model with the constant factor is closer to an equilibrium model (KMT) compared to the variable factor. \citet{Lupi2018} find that, depending on the gas properties, the clumping factor can vary significantly in their simulation, from 1 to ${\sim} 10^3$, but its distribution is far from being uniform. Several authors choose not to consider a clumping factor \citep{Richings.Schaye2016b,Hu2016,Pallottini2017,Nickerson2018}.

The second option is whether  to link the SFR explicitly to the H$_2$ content. Observations show that SF is correlated to H$_2$ \citep{McKee.Ostriker2007}, but this may be because both stars and H$_2$ form in dense and cold environments and not because H$_2$ directly triggers SF \citep{Glover.Clark2012}. \citet{Gnedin.Kravtsov2011} argue that setting the SFR proportional to the H$_2$ density avoids possibly arbitrary density and temperature thresholds since H$_2$ naturally correlates with dense, cold gas \citep[also][]{Christensen2012,Tomassetti2014,Pallottini2017}. However, \citet{Richings.Schaye2016b, Hu2016, Capelo2018, Lupi2018} and \citet{Nickerson2019} maintain a SFR proportional to total gas density. \citet{Hu2016} find that SF correlates with HI-dominated cold gas better than with H$_2$, and a  significant presence of warm, non-star-forming H$_2$ gas. \citet{Lupi2018} argue that the KS law between H$_2$ and SFR arises naturally and their arbitrary link is unnecessary.

Properly modelling the gas chemistry is not only important to accurately compute gas cooling in numerical simulation but also to compute the H$_2$ abundance to compare with observed molecular KS law \citep{Bigiel2008,Bigiel2010}. This is particularly timely given the wealth of far-infrared data coming from recent \citep[e.g. the Herschel Space Observatory][low redshift]{Pilbratt2010} and current facilities such as the Atacama Large Millimeter/submillimeter Array \citep[ALMA -][]{ALMAPartnership2015}. A more sophisticated treatment of these physical processes in numerical simulation will contribute to explore the nature of the relationship between SF and H$_2$, to test SF recipes, and to describe the H$_2$ impact on the ISM characteristics. 

In this work, we present a novel interface between an updated version of {\small GADGET-3}  \citep{Springel2005b,Beck2016} and the chemistry package {\small KROME} \citep{Grassi2014}. We explicitly follow nine primordial (H, H$_2$, He and their ions) and seven metal species and model the corresponding non-equilibrium cooling using {\small KROME}.
An external radiation field is adopted following the scheme of \citet{Haardt.Madau2012}. Additionally, a stellar radiation feedback is modelled \citep{Bruzual.Charlot2003,Lupi2018,Plat2019}. Chemical and energy feedback by Type II (SNII) and Type Ia (SNIa) Supernovae are followed within  a multiphase ISM \citep{Scannapieco2005,Scannapieco2006}. {\small KROME} has been consistently coupled to the latter scheme. 
With {\small GADGET3-K} we explore different models of H$_2$ formation on dust and investigate the impact of metallicity in the H$_2$ networks and how it affects the properties of the H$_2$ distribution and the ISM.
For this purpose, an initial condition (IC) corresponding to an idealised gas-rich Milky Way mass-size galaxy set to evolve passively is used. The initial chemical abundances  follow a radial profile with a negative. We run  this IC with different combination of parameters, dust model, SF algorithm and metallicity dependence to explore the effects of H$_2$ production and identify which models and parameters better reproduce observations.

This paper is organised as follows. In Section~\ref{sec:2}, an overview of the adopted version of {\small GADGET-3} and how {\small KROME} is grafting within it are provided. In Section~\ref{sec:3}, idealised simulations are analysed to test our method and compare the results with  observations. Additionally, we test the robustness of our method against numerical resolution. Finally, in Section~\ref{sec:4}, we summarise our findings and provide future directions for our current work.

\section{Simulation code} \label{sec:2}

In this section, we describe  {\small P-GADGET3-K}, which is the outcome of grafting the publicly available chemistry package {\small KROME} developed by \citet{Grassi2014},  into the update implementation of {\small GADGET-3} by \citet{Beck2016}. Additionally, we discuss different implementations of the SF algorithm tested in {\small P-GADGET3-K} .

\subsection{Main characteristics of the hydrodynamic code: {\small GADGET-3}}

We use a version of {\small GADGET-3} that incorporates the improved implementations of \citet{Beck2016} for the hydrodynamical evolution based on the Smoothed Hydrodynamical Particle (SPH) technique.
The density is estimated in a classic fashion from the mass distribution of gas particles weighted by the kernel function while the hydrodynamical forces are calculated according to the density-entropy formulation. However, instead of using the traditional cubic spline function, this version use the Wendland kernels \citep{Dehnen.Aly2012} with 200 neighbours, achieving better numerical convergence. As a result a more accurate density and density gradients estimations are obtained. 

Additionally, the version of {\small GADGET-3} used includes a high-order gradient computation from the full-velocity gradient matrix \citep{Cullen.Dehnen2010,Hu2014} and a shear flow limiter \citep{Balsara1995}.
It also implements a prescription for locally adaptive time-dependent artificial conduction \citep{Price2008,Tricco.Price2013,Arth2019} to treat discontinuities in the internal energy (for purely numerical reasons), which corrects for gravitationally induced pressure gradients and improves the performance of the SPH scheme in capturing the development of gas-dynamical instabilities. 

\subsection{Chemical and energy feedback and the multiphase ISM model}

Our current version, {\small P-GADGET3-K}, includes chemical and energy feedback by SNII and SNIa. For the energy SN feedback the model of \citet{Scannapieco2006} is adopted. This model has been proven to successfully trigger galactic mass-loaded winds without introducing mass-scale parameters or kicking gas particles, or suppressing the cooling in surrounding gas particles. The impact of galactic winds has been reported to naturally adapt to the potential well of the galaxy where SF takes place. A Salpeter Initial Mass Function (IMF) is adopted \citep{Salpeter1955}. The SN feedback is grafted within a multiphase model for the ISM described in detail by \citet{Scannapieco2005}. This multiphase scheme allows the coexistence and material exchange between the hot, diffuse phase and the cold, dense gas phase. The thermodynamical and chemical changes are introduced on particle-by-particle basis, considering the physical characteristics of its surrounding gas medium \citep{Marri.White2003}.

Briefly, in the SN feedback model, stars form in dense and cold gas clouds and part of them ends their lives as SNae events, injecting energy and chemical elements into the ISM. Both chemical elements and energy are distributed between the cold and hot phases of a given star particle \citep[][]{Scannapieco2008}. We assume that 70 per cent of chemical material \citep{Tissera2016} and 50 per cent \citep{Scannapieco2005} of the SN energy released are injected into the cold phase surrounding the stellar progenitor (the rest of the chemical elements and energy are injected into the surrounding hot phase). 
The released SN energy is thermalised instantaneously in the hot phase. Conversely, the cold phase stores the injected energy in a reservoir until there is  enough energy to change the entropy of the cold gas particle so that it joins its hot phase. Hence, it is a self-regulated process and does not involve sudden changes in kinetic energy. The mass-loaded galactic winds induced by the SN model of \citet{Scannapieco2008} self-regulate according to the potential wells of the galaxies.
 
We use the chemical evolution model developed by \citet{Mosconi2001}. This model considers the enrichment by SNII and SNIa adopting the elements yield prescriptions of chemical elements of \citet{Woosley.Weaver1995} and \citet{Iwamoto1999}, respectively. A set of 13 isotopes (H, $^{4}$He, $^{12}$C, $^{16}$O, $^{24}$Mg, $^{28}$Si, $^{56}$Fe, $^{14}$N, $^{20}$Ne, $^{32}$S, $^{40}$Ca and $^{62}$Zn) is followed in time.
Galactic winds are responsible for transporting metals out of the galaxies from the cold ISM into the hot circumgalactic medium. Testing different values for the injection of chemical elements, \citet{Tissera2016} found that the selected value provides a good description of the metallicity gradients of the stellar populations in the disc components of the galaxies when compared with the observational results from the CALIFA survey \citep{Sanchez-Blazquez2014}. The metallicity gradients of the gas-phase discs are also within the observed range \citep{Tissera2016a}.

The lifetimes for SNIa are randomly selected within the range [0.1, 1] Gyr. This model, albeit simple, is able to reproduce well mean chemical trends. 
Indeed, \citet{Jimenez2014,Jimenez2015}, compared the median chemical abundances generated by this SNIa lifetime model and the Single Degenerated scenario  for the delay time-life distribution \citep{Matteucci.Recchi2001}, finding similar averaged trends (within the estimated dispersion).

This multiphase scheme improves the description of the thermodynamical  characteristics of the regions where H$_2$ can be formed as it can better reproduce the cold and high-density regions, separated from the low-density ones. Without this model, the density and temperature (entropy) tend to have more homogeneous distributions towards higher and lower values, respectively \citep[][]{Marri.White2003, Scannapieco2006}. Hence, the areas where H$_2$ can be formed would not be well determined. In addition, chemical elements play a crucial role in determining the cooling rates and the dust-to-gas fractions. Having the enrichment of the baryons self-consistently described as the systems evolve \citep[see][]{Scannapieco2005, Scannapieco2008} allows us to better simulate the evolution of the H$_2$.
 
\subsection{Main characteristics of the chemistry package {\small KROME}}

The chemistry package {\small KROME} \citep{Grassi2014} is a flexible Python code that, given a set of chemical reactions, writes the necessary and optimised Fortran subroutines to solve the associated system of ordinary differential equations (ODEs) and follow the time-dependent evolution of the chemical species as well as the gas temperature. {\small KROME} employs the implicit high-order ODE solver 'DLSODES' \citep{Hindmarsh1983} to integrate this system and also provides a variety of models to describe several physical processes, including radiative and thermochemical cooling/heating, photochemistry, and dust physics. Finally, these subroutines can be embedded as a library in any hydrodynamics code. 
It has been successfully employed to study a variety of problems (environments and processes) as: black holes \citep{Latif2019}, SF activity \citep{Lupi2018, Lupi.Bovino2020}, metal and molecular chemistry in galaxies and filaments \citep{Bovino2016, Capelo2018, Seifried2017}, first-stars \citep{Latif.Schleicher2015, Piyush2019}, etc. 

\subsection{Modelling the gas chemistry and thermal processes} \label{sec:model}

Chemistry and radiative cooling/heating are computed with {\small KROME}, following the model 1a described in \citet{Bovino2016}. This model takes into account the photoheating, H$_2$ UV pumping, Compton cooling, photoelectric heating, atomic cooling, H$_2$ cooling, and chemical heating and cooling. 
Our own implementation includes the non-equilibrium rate equations for nine primordial species (i.e. HI, HII, H$_2$, H$_2^+$, H$^-$, HeI, HeII, HeIII, and e-) and also seven metal species (CI, CII, OI, OII, SiI, SiII, and SiIII). In the latter case, a linear system for the individual metal excitation levels is solved on-the-fly for the most important coolants in the ISM \citep[see][]{Glover.Jappsen2007,Maio2007}.

In addition, following \citet{Lupi2018}, we include three-body reactions involving H$_2$ \citep{Glover.Abel2008,Forrey2013}, H$_2^+$ collisional dissociation by H, and H$^-$ collisional detachment by He \citep{Glover.Savin2009}. We clarify that these reactions are taken into account only for completeness, due to the very small reaction rate expected in the density range achieved in the simulations analysed in this work. H$^-$ photodetachment is also incorporated, though its effect is expected to be relevant in low-metallicity regions. Our chemical network covers a total of 16 species and 70 selected reactions. 

The dissociation of H$_2$ occurs via two main mechanisms, the excitation to the vibrational continuum of an excited electronic state and the Solomon process when dissociating and ionising radiations are present \citep[e.g.][]{Bovino2016}.
To model this, we consider the contributions of ionising fluxes from local stellar sources (see sec.\ref{sec:sr}), and from a uniform extragalactic ultra-violet (UV) background adopting the model of \citet{Haardt.Madau2012}. Following \citet{Lupi2018}, we define ten energy bins, ranging from 0.75 to 1000 eV, to cover principally the characteristic energies for H$^-$ photodetachment, H$_2$ and H$_2^+$ dissociation, and the different hydrogen and helium ionisation states \citep{Katz2017}.

The extragalactic background is unable to reach the innermost parts of a galaxy because of the shielding by the intervening gas and dust. To account for the attenuation of the radiation in these high-density regions, we assume that the UV background is diminished by a factor $\exp(-\tau)$, where $\tau_{\mathrm{bin}} = \sum_{i} \sigma_{i,\mathrm{bin}} N_{i}$ is the optical depth for every energy bin, $\sigma_{i,\mathrm{bin}}$ is the cross section in the bin for the i-th species (pre-computed by {\small KROME}), and $N_{i}$ is the i-th species column density. 
We simplify this following \citet{Capelo2018} and assuming that the neighbouring gas surrounding each gas particle has similar properties to the given particle. Then, the column density can be estimated as $N_{i} \sim n_{i} \lambda$, where $n_{i}$ is the i-th species number density and $\lambda$ is the absorption length, which is assumed to be equal to the Jeans length $L_{J}$ of a given gas particle. \citet{Safranek-Shrader2017} show that the Jeans length is a better approximation compared to gradient-based shielding lengths.
Additionally, at high densities, gas clouds also self-shield from radiation. To consider this process and properly track the abundances in these regions without overestimating the effective radiative flux, self-shielding by H$_2$ and dust is adopted. Both terms are implemented in {\small KROME} as a sub-grid recipe following \citet{Richings2014b}.

{GADGET3-K} includes the formation of H$_2$ by both possible channels: in  gas-phase \citep[e.g. reactions 7-8 and 9-10 in ][]{Bovino2016} and by catalysis on dust grains \citep{Hollenbach.McKee1979}. The later channel dominates for high levels of metal enrichment ($Z \gtrsim 10^{-2} Z_{\sun}$) and dust-to-gas ratio.

\subsubsection{H$_2$ formation on dust}

The presence of dust is a critical aspect for the formation of H$_2$ in most of the relevant regions where SF can take place. 
We employ two models to describe this process:
\begin{itemize}
\item
The scheme of \citet[][J75]{Jura1975} that assumes fixed dust properties and no dependence on gas temperature, yielding the following rate
\begin{equation}
\frac{\mathrm{d}n_{\mathrm{H_2}}}{\mathrm{d}t} = 3 \times 10^{-17} n_{\mathrm{HI}} n_{\mathrm{H}} \frac{Z}{Z_{\odot}} C_{\rho}\ \ \mathrm{cm}^{-3} \mathrm{s}^{-1}.
\end{equation}

\item The model described by \citet{Tomassetti2014}, following the precursors schemes of \citet{Tielens.Hollenbach1985} and \citet{Cazaux.Spaans2004}, that adopts a dependence on the gas temperature resulting in a rate
\begin{equation}
\frac{\mathrm{d}n_{\mathrm{H_2}}}{\mathrm{d}t} = 3.025 \times 10^{-17} n_{\mathrm{HI}} n_{\mathrm{H}} S_{\mathrm{H}} \sqrt{\frac{T}{100\mathrm{K}}} \frac{Z}{Z_{\odot}} C_{\rho}\ \  \mathrm{cm}^{-3} \mathrm{s}^{-1},
\end{equation}
where 
\begin{equation*}
S_{\mathrm{H}} = \left[ 1 + 0.4 \left(\frac{T+T_{d}}{100\mathrm{K}}\right)^{1/2} + 0.2 \frac{T}{100\mathrm{K}} + 0.08 \left(\frac{T}{100\mathrm{K}}\right)^{2} \right]^{-1}
\end{equation*}
is the sticking probability coefficient of the H atoms, which depends on gas and dust temperatures. For simplicity, here we adopt a constant dust temperature $T_{d} = 10$ K, value typical of molecular clouds, following the works of  \citet{Tomassetti2014}, \citet{ Richings2014a} and \citet{Bovino2016}. These last two articles showed that, in the range of gas densities considered in their simulations, similar in resolution to our experiments, $T_{d}$ is not strongly affected by the interaction between dust and gas, and therefore, the approach where $T_{d}$ is calculated by solving the thermal equilibrium equation produces equivalent results to that where $T_{d}$ is kept constant. 
\end{itemize}

In these models, we do not compute the dust temperature nor the gas cooling by dust. However, its impact is not relevant in the density range  $n_{\mathrm{g}} \lesssim 10^{7} \mathrm{cm^{-3}}$  \citep[see][]{Bovino2016,Grassi2017}. This density threshold is larger than the maximum density reached in our simulations.

For both models $n_{\mathrm{H}} = n_{\mathrm{HI}} + n_{\mathrm{HII}} + n_{\mathrm{H^-}} + 2n_{\mathrm{H_2}} + 2n_{\mathrm{H_2^+}}$ is the total H number density, and we assume that the mass dust-to-gas ratio, $D \equiv \rho_{d}/\rho_{g}$, scales linearly with metallicity as $D=D_{\odot} Z/Z_{\odot}$, where $D_{\odot}=0.00934$ \citep{Yamasawa2011}.
However we note that \citet{Aoyama2017} claim that, at high redshift $(z \gtrsim 3.5)$, this correlation breakdown. This change might be important to test in cosmological simulations.

$C_{\rho}$ is a clumping factor that takes into account the possible missed H$_2$ formation in the high-density regions due to limited numerical resolution.  In the next section, we will explore the impact of the clumping factor by running the same IC  with  $C_{\rho}=1$ (i.e. no clumping factor), 10 and 100. We will also explore a variable clumping factor following \citet{Lupi2018}:
\begin{equation} \label{ec:c_var}
    C_{\rho} = 1 + b^2 \mathcal{M}^2, 
\end{equation}
where $\mathcal{M}$ is the Mach number, and $b$ a turbulent parameter which describes the ratio between compressive and solenoidal turbulence and its typical value is $b = 0.4$, for a statistical mixture of these turbulences \citep{Federrath2010b}.

\subsubsection{Thermal processes}

The thermal behaviour (cooling/heating) of the gas is an extremely important physical process that has profound effects on the evolution of a galaxy, since it has direct impact on the SF, among other  phenomena. In this work, we include via {\small KROME} the following thermal processes: HI, HeI, and HeII collisional excitation and ionisation, HII, HeII, and HeIII recombination, HeI dielectronic recombination, Compton cooling from the CMB-relevant for cosmological simulations, Bremsstrahlung \citep{Cen1992}, H$_2$ roto-vibrational cooling \citep{Bovino2016,Glover.Abel2008,Glover2015} and collisional dissociation \citep{Omukai2000}, dust surface recombination \citep{Bakes.Tielens1994}, and metal line cooling (see e.g. \citet{Grassi2014} and \citet{Bovino2016} for a thorough explanation).

Metal-line cooling is a dominant process for gas with relatively high metallicity and low density. However, at high temperatures, given the large number of metal species, their ionisation states, transitions, and collisional processes, it becomes computationally expensive to properly follow the detailed metal network and its related non-equilibrium cooling. Moreover, for $T \gtrsim 10^{4}$ K, it is reasonable to assume thermochemical equilibrium and employ photoionisation equilibrium (PIE) metal cooling tables, due to the presence of a radiation background. Therefore, for $T \geqq 10^{4}$ K, we adopt the PIE table computed by \citet{Shen2013} using Cloudy \citep{Ferland1998} and considering an extragalactic radiation background by \citet{Haardt.Madau2012}. This table provides the metal cooling rates as a function of gas temperature ($10 \leq T \leq 10^{9}$ K), density ($10^{-9} \leq n_{\mathrm{H}} \leq 10^{4}$ cm$^{-3}$), and redshift ($0 \leq z \leq 15.1$). Moreover, these rates are provided assuming solar abundances and then linearly re-scaled with the metallicity $Z$ of each particle, which is followed as a passive scalar.

On the other hand, for a gas temperature $T < 10^{4}$ K, where the assumption of equilibrium does not hold, we use the non-equilibrium cooling rates by CI, CII, OI, OII, SiI, and SiII because these elements are the most important metal coolants in the ISM \citep[see][]{Wolfire2003,Shen2010,Richings2014a}. In this case, {\small KROME} solves an equilibrium linear system using on-the-fly coolant and collider(s) abundances and temperature for the individual excitation levels of these metal atoms and ions.

\subsection{Star formation scheme} \label{sec:sf}

The processes involved in the transformation of gas into stars are complex and act on smaller scales than those involved in the galaxy assembly. To model these phenomena, SF sub-grid recipes are used. 
SF algorithms usually adopt thresholds to define the dense and cold gas that is in condition to be transformed into stars following a probabilistic approach \citep{Springel.Hernquist2003}.
The new  star particles, which are assumed to represent simple stellar populations, are stochastically created according to the probability derived from the SFR of each gas particle. The likelihood for a gas particle to be transformed into stars is computed using different schemes. In this paper, we consider two methods.

On one hand, the 'classic' recipe assumes that a gas particle with temperature $T < T_{\mathrm SF} = 10^{3}$ K and density $\rho_{\mathrm g} > \rho_{\mathrm SF} = 8\ m_{\mathrm{H}}$ cm$^{-3}$, is in condition to form a star particle according to 
\begin{equation}
p_1 = (m_{\mathrm{g}}/m_{\mathrm{s}})\left[1 - e^{-\left(\epsilon_{\ast} \mathrm{d}t/t_{\mathrm{dyn}}\right)}\right],
\end{equation}
where $m_{\mathrm{g}}$, $\mathrm{d}t$ and $t_{\mathrm{dyn}} = \sqrt{3 \pi / (32 G \rho)}$ are the mass, time step and dynamical time of each  star-forming gas particle, and $G$ is the gravitational constant. $m_{\mathrm{s}}$ will be the stellar mass created (a preset fraction $f_{s}$ of the gas particle mass) and $\epsilon_{\ast}$ is the SF efficiency parameter \citep{Springel.Hernquist2003, Krumholz.McKee2005}. 

On the other hand, following \citet{Hopkins2018}, the second method selects a gas particle to form stars based on the so-called virial parameter $\alpha$, which restricts the formation of stars only to self-gravitational regions, i.e. those particles that are prone to gravitational collapse \citep{Semenov2016}:
\begin{equation}
\alpha_i = \frac{||\nabla \otimes \mathbf{v}_i||^2 + (c_{\mathrm{s},i}/\delta r_i)^2}{8 \pi G \rho_i}.
\end{equation}

This 'self-gravity' method allows only gas particles with $\alpha < 1$ to be transformed into stars, which indicates that these gas particles cannot overcome gravitational collapse through  kinetic or thermal support. Here, $\mathbf{v}_i$ is the gas velocity, $c_{\mathrm{s},i}$ the sound speed, $\delta_i$ the particle size and $||\nabla \otimes \mathbf{v}_i||^2$ is the Frobenius norm of the velocity dispersion matrix, defined as
\begin{equation}
||\nabla \otimes \mathbf{v}_i||^2 = \sum_{i,j} \left(\frac{\partial v_i}{\partial x_j}\right)^2.
\end{equation}

We also incorporated a simple H$_2$-dependent SF algorithm that uses the fraction of molecular hydrogen $f_{\mathrm{H_2}}$ per gas particle to weight the probability according to
\begin{equation}
p_2 = 
\begin{cases}
        p_1\ f_{\mathrm{H_2}}      & Z/Z_{\odot} > 10^{-3} \\
        p_1\ (1-\alpha)  & Z/Z_{\odot} \leq 10^{-3}
\end{cases}
\end{equation}

The lower metallicity threshold adopted to weight by the H$_2$ fraction is imposed considering that in the case of pristine gas $f_{\mathrm{H_2}} \sim 0$ and hence, no SF would occur.
Previous theoretical works \citep[e.g.][]{Maio2010} and observations of very metal-poor stars \citep[e.g.][]{Frebel2010} suggest that at the metallicity $(Z_{\mathrm{crit}})$ at which the metal cooling function dominates over the molecular one, the modalities of SF transition from Population III SF regime to the standard Population II-I regime.
Different studies suggest $Z_{\mathrm{crit}}$ varies from ${\sim} 10^{-6} Z_{\odot}$ \citep{Schneider2006} to ${\sim} 10^{-3} Z_{\odot}$ \citep{Bromm.Loeb2003, Bovino2014}. For this analysis we adopt the latter as our metallicity threshold.

Both probability functions, $p_1$ and $p_2$, imply that, on average, $\mathrm{d}M_{\ast}/\mathrm{d}t = \epsilon_{\ast} m_{\mathrm g}/t_{\mathrm dyn}$, effectively ensuring that SF  follows the slope of the KS law \citep[][]{Schmidt1959,Schmidt1963,Kennicutt1989,Kennicutt1998} between SFR and gas surface densities, whereas the normalisation can be matched by tuning the SF efficiency parameter, $\epsilon_{\ast}$. We run our simulations with $\epsilon_{\ast} = 0.01$ \citep{Krumholz2012}. With this value the normalisation of the KS law is fairly well-reproduced and is consistent with the average SF efficiency computed as a function of the gravo-turbulent state of the gas in \citet{Lupi2018}.

\subsection{Stellar radiation scheme} \label{sec:sr}

In addition to the extragalactic UV background, we also include a phenomenological  model to take into account the radiation field from the nearby stellar populations. Due to the strong ionising and dissociating flux produced principally by young stars, this form of radiative feedback has a significant impact on  the H$_2$ abundance.   To estimate the contribution of the young stellar populations, we assume that every star particle generated in our simulations forms in an instantaneous SF burst, which produces a simple stellar population. A Salpeter IMF, with lower and upper mass cutoffs of $\mathrm{m_L = 0.1\ M_{\odot}}$ and $\mathrm{m_U = 100\ M_{\odot}}$, is adopted to estimate the number  of stars of a given stellar mass.
Then, we use the Salpeter IMF version of the C\&B SSP models described in \citet{Plat2019} to follow the age and metallicity dependent spectral evolution of the luminosity $L_{\ast}$ of each star particle. 
The C\&B models are an updated version of the \citet{Bruzual.Charlot2003} models, which incorporate, most notably, a new prescription for the evolution of massive stars from \citet{Chen2015} and higher-resolution spectral libraries (see \citet{Plat2019} for details).

To calculate the total flux that radiates to a gas particle, different models have been proposed. \citet{Christensen2012} calculate the average flow of nearby stars using the tree structure employed for gravity computation. Similarly, \citet{Hopkins2014, Hopkins2018} and \citet{Lupi2018} use the total luminosity and the luminosity-weighted centre of the stellar sources within each tree node. However, efficient H$_2$ and dust shielding at high densities makes the H$_2$ fraction relatively insensitive to the amount of dissociation radiation incident on molecular clouds \citep{Gnedin2009, Krumholz2009, MacLow.Glover2012}.
To consider this attenuation and shielding, \citet{Christensen2012} assume that the column length is equal to the smoothing length $h$ of the target gas particle. However, this scheme may be susceptible to resolution effects. In this regard, \citet{Safranek-Shrader2017} evaluate several common approximations for the shield length, such as the Sobolev approximation, the Jeans length, a single cell and power law approximations, and a length based on the local density and its gradient. They compare these implementations to a detailed ray tracing solution for the radiative transfer problem and find that a temperature-limited Jeans length work well to equal effective visual extinction and performs better than other local models in calculating the weighted mass abundances of H$_2$ and CO. \citet{Byrne2019} reinforce these conclusions.

Based on these results, we opt for a model that loops over gas particles gathering all star particles within a maximum distance $R_{\mathrm{c}}$, assumed to be equal to the Jeans length of the target gas particles
\begin{equation} \label{ec:tau_rmax}
L_{J\mathrm{(T<40K)}} = \sqrt{\frac{\pi c_s^{2}}{G \rho}} = \sqrt{\frac{\pi}{G \rho} \frac{\gamma T k_{\mathrm{B}}}{\mu m_{\mathrm{H}}}},
\end{equation}
where $T$ is the gas temperature with a ceiling value of 40 K, $\gamma$ and $\mu$, the adiabatic coefficient and the mean molecular weight of gas,respectively, and $k_{\mathrm{B}}$, the Boltzmann constant. 
This choice is consistent with the attenuation model adopted by KROME. 
We also impose the condition $R_\mathrm{c} = \mathrm{max}{(R_\mathrm{c}\ ,\ 5h)}$ since a factor of five represents a good compromise between collecting radiation from large enough distances and restricting the computational overhead \citep{Lupi2018}\footnote{This upper limit, $R_\mathrm{c}= 5h$, is seldom reached, except for gas particles in extreme low density regions.}.

Hence, considering  all stellar particles  within $R_\mathrm{c}$, we compute the total radiative flux, in each of ten defined energy bins (ranging from 0.75 to 1000 eV, see sec.~\ref{sec:model}), which  reaches a target gas particle as
\begin{equation} \label{ec:star_flux}
F = \sum_j \left( \frac{L_{\ast,j}}{4 \pi d_j^2}  \exp{(-\tau_j)} \right),
\end{equation}
where $L_{*,j}$ and $d_j$ are the luminosity and the distance to the $j$-th stellar particle, respectively. Since we cannot estimate the real optical depth $\tau_j$ of the total gas located between the different stellar sources and the target gas particle, for each energy bin, $\tau_j$ is approximated by using its species abundances and an absorption length equal to $d_j$ $(\tau_{j,\mathrm{bin}} \approx \sum_{i} \sigma_{i,\mathrm{bin}} n_i d_j)$.

Additionally, we consider the attenuation from dust computed from the dust-to-gas ratio dependent  scheme "Milky Way, R$_{\mathrm{V}}$ = 3.1"  by \citet{Weingartner.Draine2001}. This is a carbonaceous-silicate grain model with a Milky Way size distribution for R$_{\mathrm{V}}$\footnote{$\mathrm{R_V \equiv A(V)/E(B-V)}$ is the ratio of visual extinction to reddening}= 3.1. This value successfully reproduces observed interstellar extinction, scattering, and infrared emission, and is considered to be appropriate for typical diffuse HI clouds in the Milky Way \citep{Draine2003}. 

\subsection{Initial conditions} \label{sec:ic}

\begin{table*}
	\centering
	\begin{tabular*}{0.85\textwidth}{@{\extracolsep{\fill}} llclcccc}
		\hline
		& Run$^{(1)}$ & H$_2$-Dust model$^{(2)}$ & $C_{\rho}^{(3)}$ & SF recipe$^{(4)}$ & Metal cooling$^{(5)}$ & $Z^{(6)}$ & Stellar Rad$^{(7)}$\\
		\hline
		& c1    & J75 & 1   & Thresholds                      & non-eq & $Z_{\mathrm{g}}$     & no \\
		& c10   & J75 & 10  & Thresholds                      & non-eq & $Z_{\mathrm{g}}$     & no \\
		& c100  & J75 & 100 & Thresholds                      & non-eq & $Z_{\mathrm{g}}$     & no \\
		& cv    & J75 & Var & Thresholds                      & non-eq & $Z_{\mathrm{g}}$     & no \\
		& c100T & T14 & 100 & Thresholds                      & non-eq & $Z_{\mathrm{g}}$     & no \\
		& S     & J75 & 100 & Self-grav                       & non-eq & $Z_{\mathrm{g}}$     & no \\
		& SH2   & J75 & 100 & Self-grav \& f$_{\mathrm{H_2}}$ & non-eq & $Z_{\mathrm{g}}$     & no \\
		& SH2\_SR & J75 & 100 & Self-grav \& f$_{\mathrm{H_2}}$ & non-eq & $Z_{\mathrm{g}}$     & yes \\
		& eZ    & J75 & 100 & Self-grav \& f$_{\mathrm{H_2}}$ & eq     & $Z_{\mathrm{g}}$     & no \\
		& eZ01  & J75 & 100 & Self-grav \& f$_{\mathrm{H_2}}$ & eq     & $0.1 Z_{\odot}$ & no \\
		& eZ05  & J75 & 100 & Self-grav \& f$_{\mathrm{H_2}}$ & eq     & $0.5 Z_{\odot}$ & no \\ 
		\hline
		& LR    & J75 & 100 & Self-grav \& f$_{\mathrm{H_2}}$ & non-eq & $Z_{\mathrm{p}}$     & no \\
		& HR    & J75 & 10  & Self-grav \& f$_{\mathrm{H_2}}$ & non-eq & $Z_{\mathrm{p}}$     & no \\
		\hline
	\end{tabular*}
	\caption{Main  parameters of simulations: 
	(1) Run name. 
	(2) Model of H$_2$ formation on dust. 
	(3) Clumping factor: fixed or variable. 
	(4) SF recipe. 
	(5) Metal cooling model for the gas at low temperatures $T < 10^{4}$ K. 
	(6) Gas metallicity:
	    fixed values without further enrichment by any source, and initial primordial gas $(Z_{\mathrm{p}})$ or an initial metallicity gradient corresponding to $z {\sim} 2$ [\citep{Maiolino2008} - $Z_{g}$]. In the latter two cases, chemical evolution is fully included (see sec.~\ref{sec:2}).
	(7) Stellar radiation feedback}
	\label{tab:runs_table}
\end{table*}

We perform different simulations of the same IC representing an  isolated galaxy to test the main parameters associated with {\small KROME} and the models for the dust and SFR implementations in our code as summarised in Table~\ref{tab:runs_table}.

The  IC we adopted for the tests (hereafter, standard IC) corresponds to a disc galaxy formed by a dark matter halo of $9 \times 10^{11}$ M$_{\odot}$, a bulge of $1.30 \times 10^{10}$ M$_{\odot}$ and a disc component of $3.90 \times 10^{10}$ M$_{\odot}$. We assume that 50\% of disc mass is in the gas-phase.
The standard IC allows us to test different parameters faster. A higher resolution version is used to test convergence.

For the standard IC, the galaxy components are resolved by $2 \times 10^5$ dark matter particles of $ 4.5 \times 10^6$ M$_{\odot}$, $6 \times 10^4$ and $4 \times 10^4$ star particles of $3.25 \times 10^5$ M$_{\odot}$, in the disc and bulge components respectively, and initially $10^5$ gas particles of $1.95 \times 10^5$ M$_{\odot}$. 
The initial dark matter profile is consistent with a NFW profile \citep{NFW1996} with a virial circular velocity of 160 kms$^{-1}$. The disc component follows an exponential profile with a scale-length of $ 3.46$ kpc while the bulge follows a Hernquist profile. The gravitational softenings adopted are $0.32$, $0.20$ and $0.16$ kpc, for the dark matter, star and gas particles, respectively.

An important aspect to take into account is that the spatial initial metallicity distribution of the gaseous disc in our standard IC has been set to have an initial radial profile with a gradient of $\nabla_{\mathrm (O/H)} = -0.1$ dex kpc$^{-1}$ and the central abundance, (O/H)$_c = 12 + \mathrm{log(O/H)} = 8.5$. The later is determined so that the IC is consistent with the mass-metallicity relation at $z {\sim} 2$ \citep{Maiolino2008}. Additionally, for consistency, we choose the model of \citet{Haardt.Madau2012} for an uniform extragalactic UV background at the same redshift ($z=2$).
With this characteristics the IC represents a massive gas-rich galaxy that evolves transforming its gas reservoir in isolation since $z {\sim} 2$. Although it is an idealised IC, it allows the exploration of the H$_2$ fraction and SF activity evolution using different model prescriptions. 

In Section \ref{sec:num_resol} the effects of numerical resolution will be evaluated. For this purpose, we run a high resolution IC which adopt the same characteristics of the standard IC but increasing the number of particles per galaxy component:  $10^6$ dark matter particles (mass of $ 9 \times 10^5$ M$_{\odot}$) , $10^5$ star particles (mass of $1.3 \times 10^5$ M$_{\odot}$) in the bulge, and $4 \times 10^5$ (mass of $4.8 \times 10^4$ M$_{\odot}$) in the disc. In addition, $10^6$ initial gas particles (initial mass of $1.9 \times 10^4$ M$_{\odot}$) are available. The respective gravitational softenings for all components are $0.2$, $0.15$, $0.1$ and $0.08$ kpc, respectively.
For this case, the initial chemical content of the gas is pristine, that is, it is initially composed of H and He only, without the imposition of any metallicity gradient (as done in some previous works). A lower resolution counterpart (with pristine gas) has been also performed in order to properly assess the impact of numerical resolution under similar conditions.

\section{Results} \label{sec:3}

In this section, we focus on assessing the impact of the clumping factor and the models of H$_2$ formation on dust (hereafter H$_2$-dust models in this text). For this purpose, we analyse the distributions of H$_2$ and HI and the implications of numerical resolution.
In addition, we explore the metal-dependence of the H$_2$ content and its impact on the  SF activity and the KS law.
The analysis is performed at $t \sim 0.6$ Gyr of evolution of the systems when the main starburst has ended, but the systems is still actively forming stars (SFR $\sim 8$ M$_\odot$ yr$^{-1}$). We stack the analysed quantities over a period of 
$0.4 \leq t \leq 0.8$ Gyr, to provide overall estimations which are more stable against possible sharp variations due to numerical resolutions.

\subsection{Genesis of H$_2$ on dust: the formation models and clumping factor}

\begin{figure*}
    \centering
	\includegraphics[trim={0 25 0 0}, width=\textwidth]{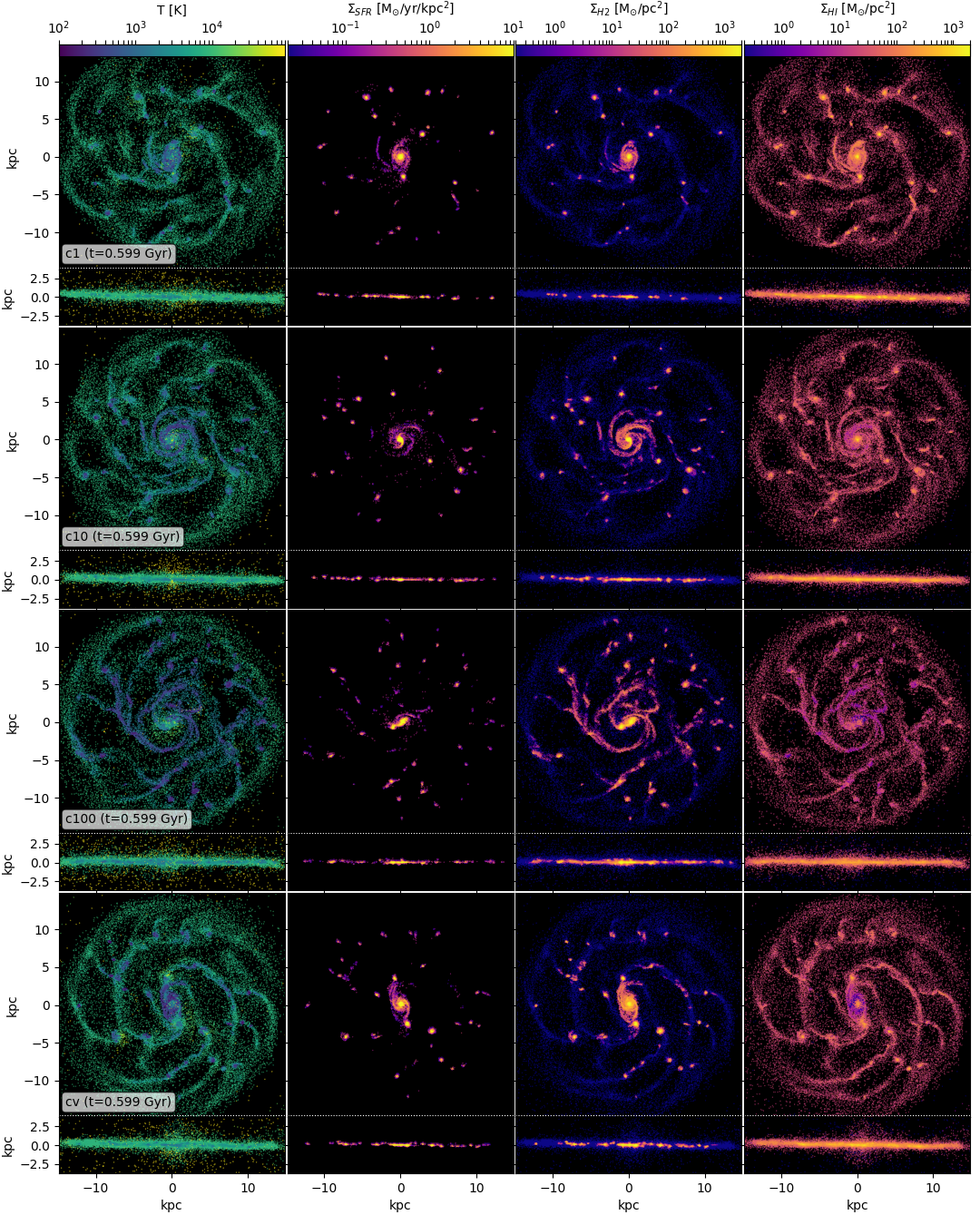}
    \caption{Face-on and edge-on maps (small panels) of the mass-weighted gas temperature (first column) and the projected surface density of SFR (second column), H$_2$ (third column) and HI (fourth) content in c1 (fist row), c10 (second row), c100 (third row) and cv (fourth row) runs of our idealised spiral galaxy.}
    \label{fig:d_colormap_1}
\end{figure*}

\begin{figure*}
    \centering
	\includegraphics[trim={0 40 0 10},width=\textwidth]{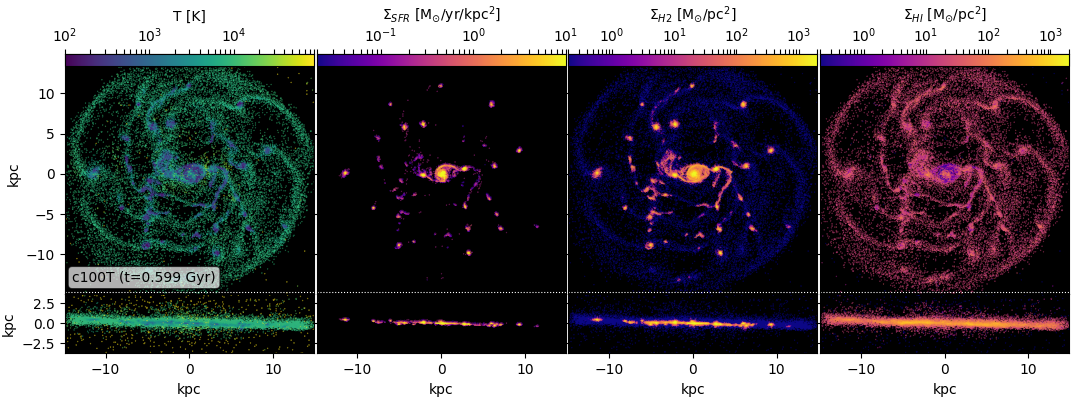}
    \caption{Face-on and edge-on maps for c100T simulations as shown in Fig.~\ref{fig:d_colormap_1}.}
    \label{fig:d_colormap_2}
\end{figure*}

\begin{figure*}
    \centering
	\includegraphics[trim={0 30 0 10},width=\textwidth]{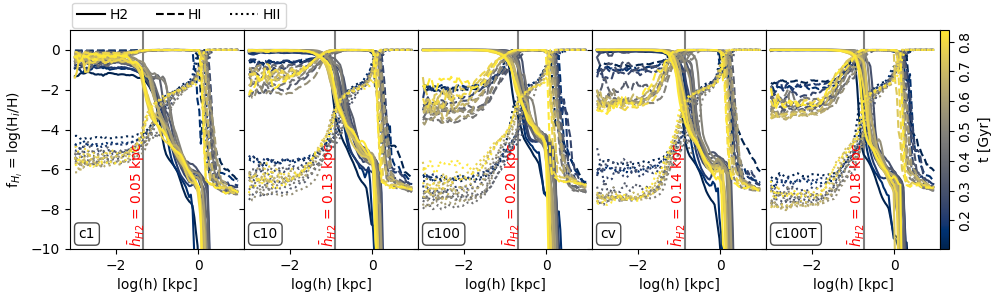}
    \caption{Vertical distribution of the fractions of H$_2$ (solid lines), HI (dashed lines) and HII (dotted lines) for simulations c1, c10, c100, cv, c100T. The vertical lines denote the average $h_{\mathrm{H_2}}$, the perpendicular distance from the middle plane of galaxy, where $f_{\mathrm{H_2}} = 0.1$. We define the thickness of the molecular discs as $2 \times h_{\mathrm{H_2}}$.}
    \label{fig:height_fH2}
\end{figure*}

\begin{figure*}
    \centering
	\includegraphics[trim={0 30 0 30},width=\textwidth]{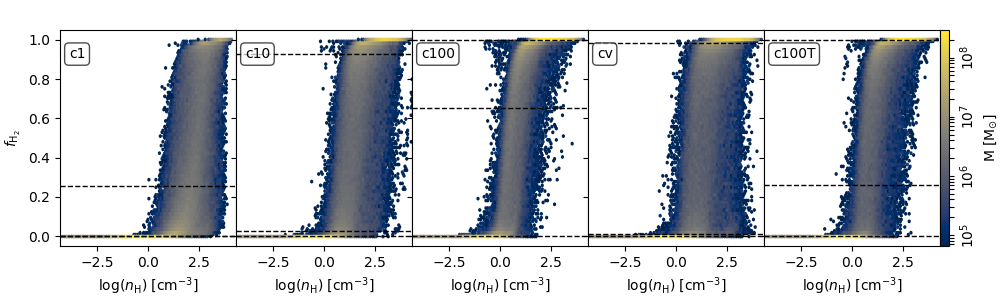}
    \caption{Phase diagrams of H$_2$ mass fraction versus $n_{\mathrm{H}}$ per particle, for runs with  c1, c10, c100, cv (a variable $C_{\rho}$) and c100T. The horizontal dashed lines indicate the  25, 50 and 75 percentiles of the distributions of H$_2$ fractions. In the leftmost panel, the 50th and 75th percentiles are very close to values of $f_{\mathrm{H_2}}<10^{-2}$.}
    \label{fig:dens_fH2}
\end{figure*}

\begin{figure*}
    \centering
	\includegraphics[trim={0 25 0 0},width=\textwidth]{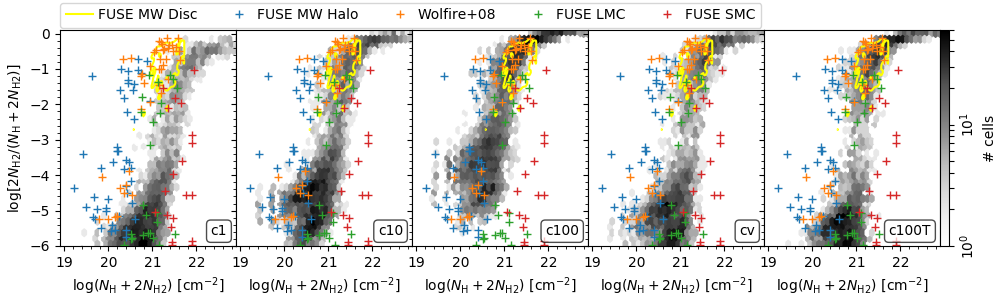}
    \caption{H$_2$ column density fractions, computed in a square grid projected onto the galactic plane, for the runs shown in Fig.~\ref{fig:dens_fH2}, compared to observations of molecular clouds in the LMC, SMC (\citealt{Tumlinson2002} - green, red cross) and MW (disc data from \citealt{Shull2021} - yellow contour enclose all of them; halo data from \citealt{Gillmon2006} - blue cross; recompiled data by \citealt{Wolfire2008} - orange cross). The colour scale show the number of cells per bin.}
    \label{fig:obs_2}
\end{figure*}

\begin{figure}
    \centering
	\includegraphics[trim={0 0 0 0},width=.47\textwidth]{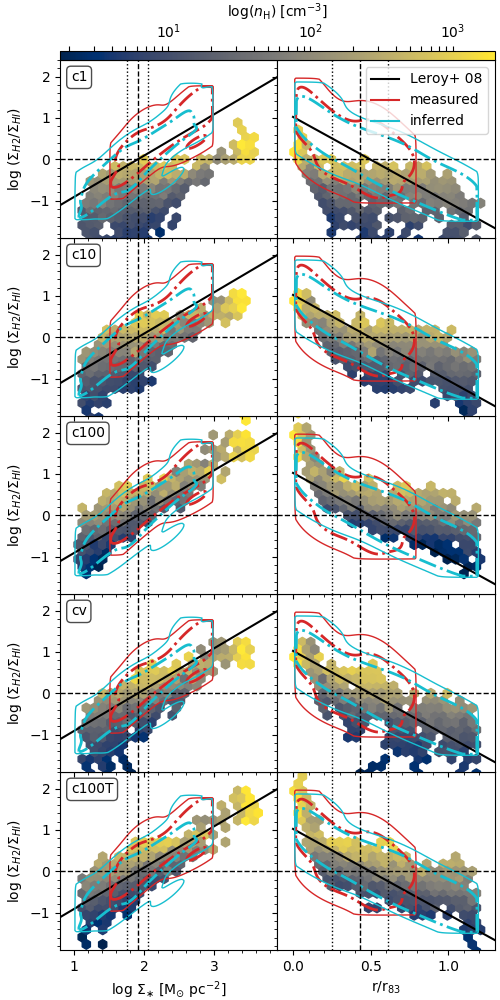}
    \caption{H$_2$-to-HI ratio, $R_{\mathrm{mol}} = \Sigma_{\mathrm{H_2}} / \Sigma_{\mathrm{HI}}$, as a function of $\Sigma_{\ast}$ (left) and radius (right). The continuous black lines are the \citet{Leroy2008} fits of $R_{\mathrm{mol}}$ and the contours enclose the 50\% and 95\% of their observational data. The red contours enclose the pixel-by-pixel measurements of $R_{\mathrm{mol}}$ in spirals and the cyan contours contain points obtained from tilted rings in spiral galaxies with $R_{\mathrm{mol}}$ inferred from $\Sigma_{\mathrm{SFR}}$ and $\Sigma_{\mathrm{HI}}$ assuming a fixed SFE(H$_2$). The horizontal dashed lines show $R_{\mathrm{mol}} = 1$. The vertical dashed lines  are the values of $\Sigma_{\ast}$ (left panels) and  radius (right panel) estimated by  \citet{Leroy2008} at which  the HI-to-H$_2$ transition takes place (the vertical - dotted lines represent the standard dispersion).}
    \label{fig:leroy_2008}
\end{figure}

In this section we analyse the J75 H$_2$-dust scheme, for different clumping factors, $C_{\rho} = 1, 10, 100$ and a variable $C_{\rho}$ ($cv$). Additionally, we perform a run with T14 H$_2$-dust model and $C_{\rho}=100$ for comparison.

First we focus on J75 H$_2$-dust scheme, for different clumping factors, $C_{\rho} = 1, 10, 100$.
In Fig.~\ref{fig:d_colormap_1} we show the face-on and edge-on projected gas density maps coloured by the mass-weighed temperature
\footnote{The face-on and edge-on temperature maps are constructed by weighted the temperature by the gas mass particle located in projected bins of ${\sim} 2\times \epsilon_{\mathrm g}$ length size.}
(first column) and the projected surface distributions of SFR  ($\Sigma_{\mathrm{SFR}}$; second column), H$_2$ ($\Sigma_{\mathrm{H_2}}$; third column) and HI ($\Sigma_{\mathrm{HI}}$; fourth column) surface densities, for runs with the J75 H$_2$-dust model and different constant $C_{\rho}$. Because the rate of H$_2$ synthesis in dust surface is directly proportional to the clumping factor, as it increases, the molecular content in the central regions grows while it extends to higher galactocentric distance and lower gas densities for higher $C_{\rho}$.

Otherwise, the spiral arm patterns in the face-on projection are very similar in all runs, suggesting that the effect on the global dynamics is not very relevant. Nonetheless, most of H$_2$ still resides in the dense regions of the galaxy spiral arms, where the gas is expected to be cold and turbulent. The HI distributions are more homogeneous as expected. However, for smaller $C_{\rho}$ they are concentrated preferentially in gas clumps located in the densest regions.
As the clumping factor increases, we observe that the gas clumps, due to favourable high densities and low temperatures, begin to be dominated by the molecular content catalysed by dust and, therefore, the HI surface density decreases whereas the  H$_2$ abundance gets more prominent.
However, these differences in H$_2$ abundance do not have a large effect on the gas density-temperature configuration because, in the densities and metallicities considered in these simulations, molecular cooling has an average impact lower than metal cooling\footnote{We note that the numerical resolution of these simulations is not high enough to follow the chemistry of CO, and therefore the cooling processes considered do not include this molecule. However, in reality, the synthesis of H$_2$ would be followed by the formation of CO, and cooling by CO rotational levels would further reduce the temperature from ${\sim} 100$ K down.}.

In the bottom row of Fig.~\ref{fig:d_colormap_1}, we show the same distributions but assuming a variable clumping factor (section~\ref{sec:model}, ec.~\ref{ec:c_var}). For the SFR density and associated molecular clumps, their distributions are clumpier because large values of $C_\rho$ \citep{Lupi2018} are generated in dense, cold gas regions compare to $C_\rho$ values outside them. This is due to its quadratic dependence on the velocity field (see section~\ref{sec:model}) that strongly favours supersonic turbulent motions, characteristic of the above clouds \citep{Larson1981}.

To assess the effects of varying the dust model in Fig.~\ref{fig:d_colormap_2} we show the face-on and face-edge distributions for the run with T14 H$_2$-dust models, which includes a dependence on the gas temperature and a fixed $C_{\rho} = 100$. As can be seen, the gas distributions are slightly smoother compared to previous, with less defined spiral patterns. In addition, the high density regions of the spiral arms exhibit a more vigorous H$_2$ formation with a more clumping distribution.

In the edge-on view (small panels in Figs.~\ref{fig:d_colormap_1} and ~\ref{fig:d_colormap_2}), we can see that runs with larger clumping factor exhibit thicker molecular discs. This can be clearly appreciated in Fig~\ref{fig:height_fH2}, where the vertical distributions of H$_2$, HI and HII fractions are shown. The fraction $f_{\mathrm{H_{i}}} = M_{\mathrm{H_{i}}}/M_{\mathrm{H}}$ is generically defined for each particle as the ratio between the mass of a species $i$ of hydrogen and the mass of the total hydrogen content. With this information we define the thickness of the molecular discs as $2 \times h_{\mathrm{H_2}}$, where $h_{\mathrm{H_2}}$ is the height from the rotation plane of the galaxy at which $f_{\mathrm{H_2}} = 0.1$. For $f_{\mathrm{H_2}} < 0.1$ we assume the discs are dominated by HI \citep{Leroy2008}. As shown in  Fig~\ref{fig:height_fH2}, we find that the average disc thickness effectively grows as  $C_{\rho}$ takes higher values, from $h_{\mathrm{disc}} \sim 100$ pc for c1, $h_{\mathrm{disc}} \sim 260$ pc for c10 and $h_{\mathrm{disc}} \sim 400$ pc for c100. We also see that in all cases the molecular disc becomes slightly thinner as the system evolves.
Adopting a variable clumping factor yields $h_{\mathrm{disc}} \sim 140$ pc similar to c10, because the median of $C_{\rho}$ distribution is ${\sim}13 \pm 4$ with 25 and 75 percentiles at $C_{\rho} = 25$ and $65$, respectively. These values are in very good agreement with those reported by \citet{Capelo2018}.

After a first qualitative comparison, we now quantitatively assess the differences amongst the models. The amount of H$_2$ varies significantly when we vary the clumping factor and/or the H$_2$-dust scheme (since it is the main formation channel for our initial conditions). Figure~\ref{fig:dens_fH2} shows the phase diagrams of H$_2$ mass fraction versus $n_{\mathrm{H}}$ (cm$^{-3}$) per gas particle, of runs c1, c10 and c100, cv and c100T. Additionally, the dotted lines indicate the percentiles 25, 50 and 75 of each H$_2$ fraction distribution. As expected, a larger clumping factor increases the amount of H$_2$ per gas particle. For c1 run $(C_{\rho} = 1)$ just over 25\% of gas mass has an $f_{\mathrm{H_2}} > 0.25$ while for c10 run $(C_{\rho} = 10)$,  45\% of gas mass has this molecular fraction and the 25\% exhibits an $f_{\mathrm{H_2}} > 0.9$. In the case of the run c100 $(C_{\rho} = 100)$ the 50\% of gas mass has a molecular fraction of $f_{\mathrm{H_2}} > 0.75$ and the 25\% display an $f_{\mathrm{H_2}} > 0.95$. With the same previous clumping factor, the T14 model (c100T) has the 50\% of gas particles with $f_{\mathrm{H_2}} > 0.52$ and the 25\%, with $f_{\mathrm{H_2}} > 0.95$.

Then, the selection of the model for H$_2$ formation on dust is relevant, but the choice of $C_{\rho}$ is more significant at least for this numerical resolution. 
These results are in agreement to those reported by \citet{Capelo2018}, where the impact of varying $C_{\rho}$ in the J75 scheme and in the model of \citet{Cazaux.Spaans2009} is analysed. As they argued, this is because their simulations, and ours, involve relatively high metallicity where the behaviour of H$_2$-dust schemes are similar.
This key aspect is well reflected in the cv run (J75 scheme with variable $C_{\rho}$) where the 25\% of gas particles  have an $f_{\mathrm{H_2}} > 0.95$ similar to run c100, but 50\% of them have an $f_{\mathrm{H_2}} < 0.1$ as in run c10.
Additionally, in Section \ref{sec:metallicity} we will discuss the effects of metallicity on the H$_2$ production and cooling.

In all runs, the transition from the neutral to the fully molecular phase occurs between   $n_{\mathrm{H}} \sim 1$ and ${\sim} 100$ cm$^{-3}$, in agreement with  previous works \citep[e.g.][]{Gnedin2009,Christensen2012}, where the gas becomes fully molecular above $n_{\mathrm{H}} {\sim} 100$ cm$^{-3}$. However, the transition region becomes thinner and moves at lower densities as the clumping factor increases for J75 model.
Conversely, a variable clumping factor produces a wider transition zone because of its dependence on dynamical properties of the gas particles. Note that the density estimations in simulations is limited by  numerical resolution, which is why this factor is incorporated into the models. Nevertheless, it has been shown that for higher numerical resolution, clumping factors are not required \citep{Nickerson2019}. The impact of resolution will be discussed in section~\ref{sec:num_resol}

We now compare the H$_2$ column density with local observations \citep{Lupi2018}. For this purpose we compute the column density of both atomic $(N_{\mathrm{H}})$ and molecular $(N_{\mathrm{H_2}})$ hydrogen, in a square grid with cells of 1 kpc size, projected onto the galactic plane. In Fig.~\ref{fig:obs_2} $N_{\mathrm{H_2}}$ as a function of the total column density, defined as $N_{\mathrm{H_2}} +N_{\mathrm{H}}$, is displayed for c1, c10, c100, c100T and cv experiments. The results are binned in 100 log-spaced bins along both axes. The colour code represents the 
number of cells in each bin.
Observational results are included for comparison, where measurements from absorption spectra of distant AGNs and nearby stars are taken primarily from the Copernicus and FUSE surveys for the Milky Way (MW) and the Large and Small Magellanic Clouds (LMC and SMC). We consider the data by \citet{Tumlinson2002} for the LMC and SMC, with $\bar{Z}_{\mathrm{LMC}} \sim 0.4 Z_{\odot}$ and $\bar{Z}_{\mathrm{SMC}} \sim 0.1 Z_{\odot}$, respectively. For the MW ($\bar{Z}_{\mathrm{MW}} \sim Z_{\odot}$), the data for the disc \citep{Shull2021} and the halo \citep{Gillmon2006} are depicted . Also we include the recompiled data by \citet{Wolfire2008}.

Additionally, in Fig.~\ref{fig:leroy_2008} we compare the projected H$_2$-to-HI surface density ratio, $R_{\mathrm{mol}} = \Sigma_{\mathrm{H_2}}/\Sigma_{\mathrm{HI}}$, with the observational data of 23 nearby disc galaxies from the HERACLES survey collected by \citet{Leroy2008}. 
These authors estimate different quantities for 23 nearby star-forming galaxies (of which 12 are large spiral galaxies). Their radial profiles are obtained within $1.2r_{25}$\footnote{The isophotal radius corresponding to 25 B-band magnitudes per square arcsec.}
using the means for $\Sigma_{\mathrm{HI}}, \Sigma_{\mathrm{H_2}}, \Sigma_{\mathrm{SFR}}$ and the median $\Sigma_{\ast}$ within $10''$ wide rings inclined. Since the average distance of this spiral galaxy subsample is ${\sim} 8.5$ Mpc, the mean width of the ring is ${\sim} 400$ pc $(10'' \sim 4.8 \times 10^{-5})$. Hence we use the same value for our radial binning within $1.2r_{83}$\footnote{$r_{83}$ is the radius that encloses 83\% of the luminous mass. For an exponential disk \citep{Freeman1970}, $r_{\mathrm{83}}$ corresponds to the de Vaucouleurs 25 B-mag arcsec$^{-2}$ photometric radius. For the simulations we assume $M/L = 1$}.
For these galaxies, \citet{Leroy2008} found that the transition between an HI and H$_2$-dominated ISM ($R_{\mathrm{mol}}=1$) is a well-determined function of locally defined quantities in the galaxy that occur at characteristic values for most of the parameters considered, such as the gas $(\Sigma_{\mathrm{gas}})$ and star $(\Sigma_{\ast})$ surface density.
Also, they found that when the ISM is mostly H$_2$-dominated in the inner parts of discs of spiral galaxies, the SF efficiency (SFE, defined as the ratio of the star formation rate to the gas mass of a parent cloud) is roughly constant as a function of the mentioned parameters ($\mathrm{SFE(H_2)} = 5.25 \pm 2.5 \times 10^{10}\ \mathrm{yr}^{-1}$ ; see section~\ref{sec:KSlaw}).

However, when the ISM is HI-dominated, the SFE declines steadily with increasing radius or other quantities covariant with it, as $\Sigma_{\ast}$, pressure or orbital time. They concluded that this implies that, while SF within GMCs is largely decoupled from environment, the synthesis of H$_2$ from HI, and by extension, the cloud formation, sensitively depends on local conditions. Thus, the behaviour of the SFE can be explained through two processes, the SF within GMCs and the formation of GMC itself. Then, the formation efficiency would prove to be the product of a constant SFE from H$_2$ and $R_{\mathrm{mol}}$, which is a function of local conditions.

In Fig.~\ref{fig:leroy_2008}, red contours enclose the 50\% (dash-dotted) and 95\% (solid) of direct measurements of $R_{\mathrm{mol}}$ from CO and HI assembled following the methodology used by \citet{Blitz.Rosolowsky2006}, while the cyan contours contain the inferred $R_{\mathrm{mol}}$ from $\Sigma_{\mathrm{SFR}}$. Here, because of the limited sensitivity of the CO data, these direct measurements of $R_{\mathrm{mol}}$ seldom probe far below 1. Therefore, \citet{Leroy2008}, assuming a fixed SFE(H$_2$) for different concentric tilted rings, convert $\Sigma_{\mathrm{SFR}}$ into $\Sigma_{\mathrm{H_2}}$ and then divide this by the observed $\Sigma_{\mathrm{HI}}$ to compute $R_{\mathrm{mol}}$ for that ring.
In addition, the continuous lines show the fits found by \citet{Leroy2008}, the horizontal dashed lines denote $R_{\mathrm{mol}} = 1$ ($\Sigma_{\mathrm{HI}} = \Sigma_{\mathrm{H_2}}$) and the vertical dashed lines show the observational estimations of $\Sigma_{\ast}$ and $r/r_{83}$ at the HI-to-H$_2$ transition (the 1-$\sigma$ uncertainties of these estimates are denoted by the vertical dotted lines).

As can be seen from this figure, the agreement between direct measurements of $R_{\mathrm{mol}}$ and the simulation estimates is quite good. $R_{\mathrm{mol}}$ is a continuous function of environment spanning from the H$_2$-dominated (${\sim} 10$) to HI-dominated (${\sim} 0.1$) ISM, from inner to outer galaxy discs, and over a wide range of ISM pressures. 

All our models generate relations within the observational range. In particular, the runs with $C_{\rho} = 100$ show the best fits to observational distributions as a function of stellar mass density and radius remarkably well, independently of the dust-formation model. The model with a variable $C_{\rho}$ (cv) exhibits a larger dispersion than c100 or c100T distributions and, on average, underestimates the observed measurements. However, as emphasised by \citet{Lupi2018}, who obtained similar results to ours, their variable clumping factor model (see section~\ref{sec:model}, ec~\ref{ec:c_var}) compensates for omitted H$_2$ formation in unresolved high-density gas regions without any calibration.

\subsection{Star formation activity}

\begin{figure}
    \centering
	\includegraphics[trim={0 0 0 0},width=.47\textwidth]{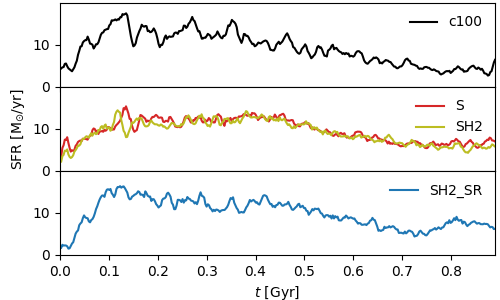}
    \caption{The SF histories of three runs with different SF algorithms: standard (so-called c100; upper panel), 'self-gravity' method (S; red) and 'self-gravity' method with the SF probability dependent on H$_2$ fraction (SH2; olive - middle panel), and the last recipe including stellar radiation feedback (SH2\_SR; lower panel). All of them have been run with J75 dust model and $C=100$}
    \label{fig:sfr}
\end{figure}

\begin{figure*}
    \centering
    \includegraphics[trim={0 25 0 0},width=\textwidth]{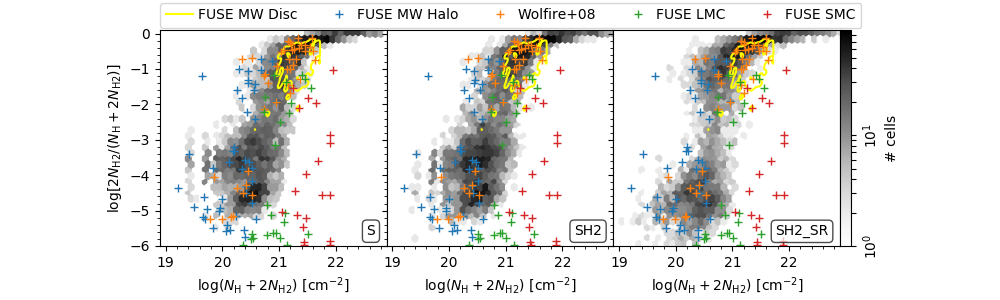}
    \caption{Observational comparisons  of H$_2$  column density fractions for runs S (standard SF), SH2 (H$_2$-dependent SF) and SH2\_SR (H$_2$-dependent SF and stellar radiative feedback). Symbol codes are given in Fig.~\ref{fig:obs_2}.}
    \label{fig:SH2_obs1}
\end{figure*}

\begin{figure}
    \centering
	\includegraphics[trim={0 10 0 20},width=.47\textwidth]{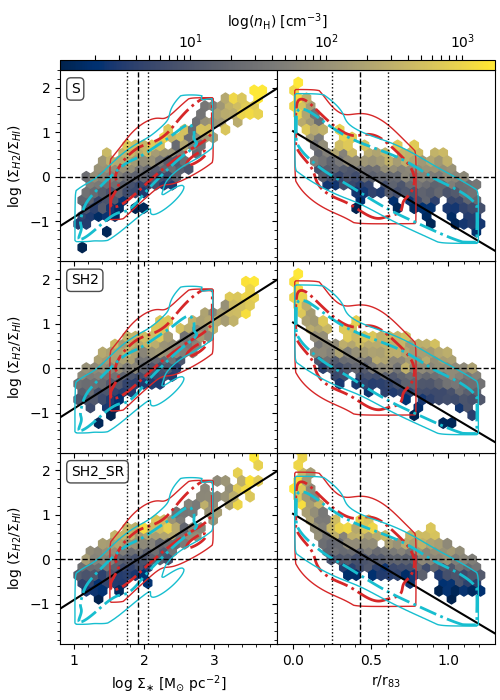}
    \caption{Observational comparisons for runs S, SH2 and SH2\_SR. H$_2$-to-HI ratio as a function of $\Sigma_{\ast}$ (left) and radius (right). See Fig.~\ref{fig:leroy_2008} for details on the contours.}
    \label{fig:SH2_obs2}
\end{figure}

\begin{figure*}
    \centering
	\includegraphics[width=\textwidth]{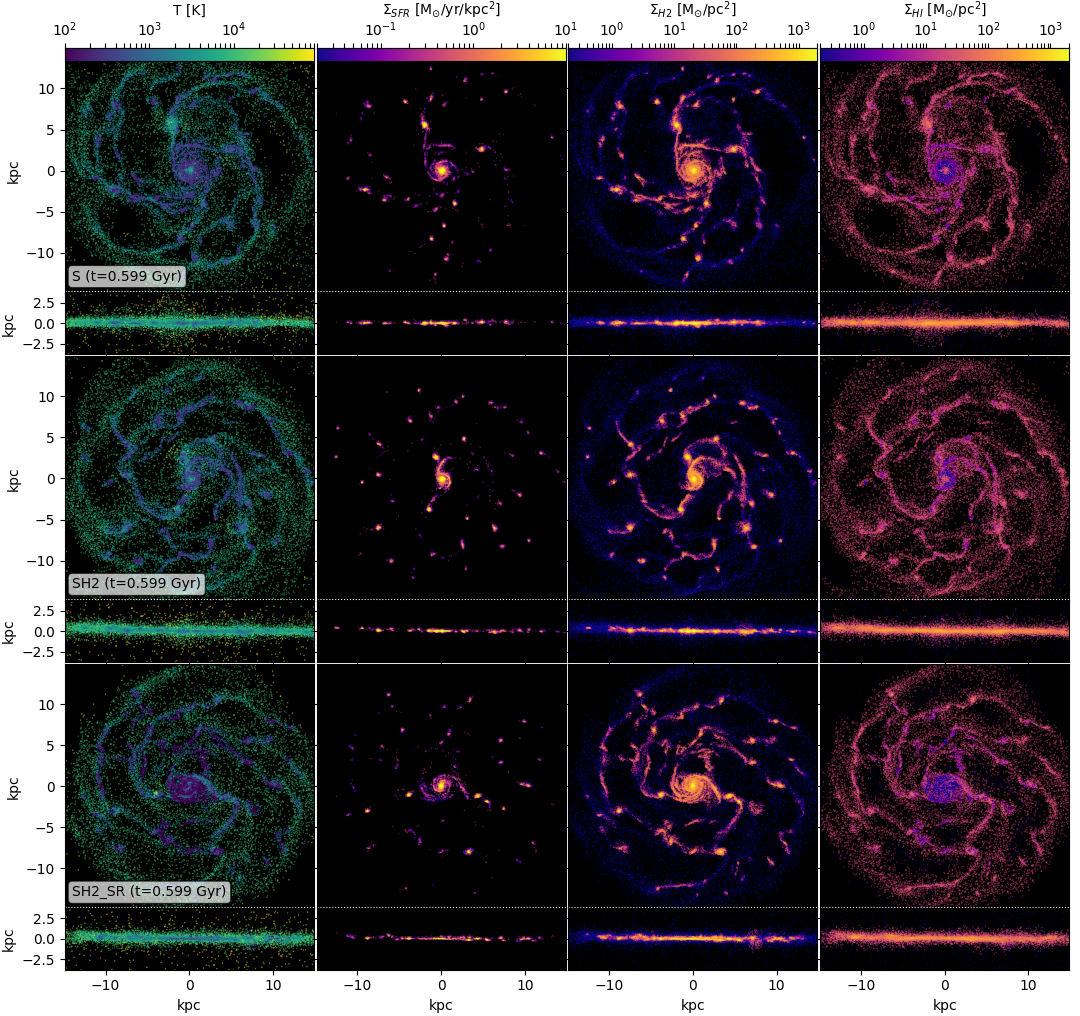}
	\vspace{-0.5cm}
	\caption{Face-on and edge-on maps of S (upper panels), SH2 (middle panels) and SH2\_SR (lower panels) simulations that adopts different SF algorithms as described in section~\ref{sec:model}. See Fig.~\ref{fig:d_colormap_1} for a detailed description.}
    \label{fig:sf_colormap}
\end{figure*}

\subsubsection{SF history}

In this section, we explore different SF schemes and assess the impact of the H$_2$  modelling. In Fig.~\ref{fig:sfr} we show the SF history of four runs, which adopted different SF algorithms: standard (so-called c100), S, SH2, SH2\_SR. They all use the J75 H$_2$-dust scheme with $C_{\rho}=100$, which best reproduces the high column density observations for this level of numerical resolution (Fig.~\ref{fig:obs_2}). As shown in Table~\ref{tab:runs_table}, the c100 run includes the standard SF model, while S and SH2 incorporate the 'self-gravity' scheme \citep{Hopkins2013b, Hopkins2015} with an independent and dependent probability on the H$_2$ fraction $(f_{\mathrm{H_2}})$, respectively. Finally, SH2\_SR  includes the last algorithm and the stellar radiation feedback scheme (see section~\ref{sec:sf} and ~\ref{sec:sr} for a detail description of these algorithms).

Our ICs assume an initial exponential disc, which quickly evolves spiral substructures. SF is high during the first 0.1 Gyr, and then relaxes as the disc settles into its semi-steady state in all runs. For the c100 run which assumes density and temperature thresholds for the SF recipe, the SFR history shows more significant oscillations with many peaks and valleys interspersed, due to the combined effects of SN feedback and adopted thresholds. 
This is because after a starburst, the release of SN energy into the nearest gas neighbours produces the heating and disruption of the cold dense gas clouds, quenching the subsequent SF activity until they dissipate the additional energy and meet the SF criteria again. In addition, since the generation of new stars takes place in dense regions, statistically this process involves particles that are located in nearby regions with similar thermodynamic properties, resulting in  relatively synchronous event that can explain the saw-tooth pattern of the SFR history in c100 run.

On the other hand, although models with self-gravity conditions show a global SFR that resembles the previous simulation, their SF activities are smoother over time so that  the gas is depleted more slowly. Consequently, in the initial stages, they have SFRs smaller than c100. Additionally, the SFRs decline less abruptly with time. Hence, after ${\sim} 0.35$ Gyr, this trend is reversed and the SFRs of simulations S and SH2 become systematically larger and more spatially extended than that of c100. We should point out that we run all simulations for 2 Gyr in order ensure that 1 Gyr is indeed a representative time of their equilibrium state.
The smooth behaviour of the SF history in SH2 is affected by the stellar radiative feedback as shown in SH2\_SR run. 
This radiation also heats and dissipates the cold and dense clouds in the ISM, and plays a major role in affecting the molecular content of the immediately contiguous gas clouds due to the significant increase of the H$_2$ dissociation rate. This effect lasts for short periods of time, corresponding to the lifetime of massive stars ($\lesssim 0.03$ Gyr), which are the main sources of FUV radiation. The lower H$_2$ fractions produce the decrease of the SFR in these regions, but the rate recovers again as the radiation from massive stars decreases over time.

\subsubsection{Comparison of H$_2$ abundances with observations}

In  Fig~\ref{fig:SH2_obs1} we show the H$_2$ column density as a function of the total column density for S, SH2 and SH2\_SR, compared to observations as in Fig.~\ref{fig:obs_2}.
The inclusion of self-gravity in the SF algorithm did not significantly modify this distribution compared to c100 (Fig.~\ref{fig:obs_2}). Similarly, weighting the SF algorithm by the H$_2$ abundances produces comparable distributions in general. 
However, SH2\_SR better reproduces the H$_2$ column density fractions for low column densities and exhibits a jump in their distribution at $\mathrm{\log[2N_{H2}/(N_H+2N_{H2})] \sim -3}$. 
This jump is probably the result of the abrupt H2 turn-on due to the effects of self-shielding \citep{Wolfire2008}. At highest values of FUV radiation, the $f_{\mathrm{H2}}$ is lower due to the rapid H2 photodissociation. As the cloud column density (and thus $\tau$) increases, the local FUV field drops due to dust extinction. In addition, the self-shielding by H2 raises the $f_{\mathrm{H2}}$ even more as the column density increases.
The treatment of radiative stellar feedback regulates the production of H$_2$ as a function of total column density reproducing fractions which are more comparable to observational data. This is in agreement with \citet{Gillmon2006} and \citet{Shull2021}, although a more detailed comparison with the MW observations is beyond the scope of this paper.

In addition, Fig.~\ref{fig:SH2_obs2} shows less dispersion in the distribution of $R_{\mathrm{mol}}$ in SH2\_SR while still reproducing the observed data. However, in the outermost regions of the simulated galaxy $(r/r_{83} \gtrsim 0.8)$ there is still an excess as in the rest of the experiments executed with constant $C_{\rho} $ (Fig.~\ref{fig:leroy_2008}). In fact, the only experiment that better reproduces this observational relation is the run with variable $C_{\rho} $. 
The objective of this parameter is to compensate for the possible formation of H$_2$ in  high-density regions which are not numerically resolved. However, a constant value is equally applied to every region, even where lower density is well resolved. Our results suggest that a variable $C_{\rho} $, which takes into account the thermodynamical properties of the gas, might be the best option \citep{Lupi2018}. We will explore this in a future work.

Nevertheless we must also consider that we are comparing our simulated galaxies that are gas-rich systems with spiral galaxies observed at $z {\sim} 0$, which have much lower total gas fractions and column densities.
So, high $R_{\mathrm{mol}}$ values at large radii, coming from very massive clumps as we can see in Figs.~\ref{fig:d_colormap_1}, \ref{fig:d_colormap_2} and \ref{fig:sf_colormap} (face-on view), could be a feature of more gas-rich and clumpy galaxies that are more frequent at $z {\sim} 2$.

The spatial distributions of the gas properties and $\Sigma_{\mathrm{SFR}}$ showed in Fig~\ref{fig:sf_colormap} are slightly different among the three experiments. The S run shows a wider and more open spiral pattern for $\Sigma_{\mathrm{HI}}$  while the SH2 and SH2\_SR runs present a tighter spiral structure where the HI is more extended. This is because most of H$_2$ content resides in the coldest and densest regions of the arms, promoting the SF activity only in these areas. Therefore, the molecular gas has shorter depletion times (defined as $\tau_{\mathrm{d}} = \Sigma_{\mathrm{SFR}}/\Sigma_{\mathrm{H_2}}$) than similar regions in the S run.
Nevertheless, in SH2\_SR the aforementioned H$_2$ spatial distributions are more dispersed than in SH2, although it shares with the latter simulation the same SF recipe (i.e. associated to molecular clouds). Furthermore, due to the dissociating action of the stellar radiation from young stars, the molecular gas regions are slightly more extended and less dense, which translates into a weaker star formation activity in these regions. 

Similarly, \citet{Byrne2019} using simulations of isolated disc galaxies and cosmological simulations of dwarf galaxies, compared a shielding-based SF model (i.e., coupling the SF probability to the amount of dust shielding) to a H$_2$-based model and to a temperature ceiling model. 
The comparison with the H$_2$-based recipe shows that the main difference is the SF at higher temperatures and lower densities found in the shielding-based scheme, which  requires higher densities than the temperature-ceiling model to have active SF, on average. As in our simulations, one consequence of the different star-forming gas densities is the presence of denser gas in the run using the H$_2$ scheme, which also leads to more pronounced gas spiral structure. This can be explained by the time delay needed to form H$_2$ prior to SF. During that time, the gas can also collapse gravitationally and reach higher densities before spawning a star particle. Models without an explicit link between H$_2$ and SF, allow earlier formation activity while the gas is at a lower density.

\subsection{The Kennicutt-Schmidt Law} \label{sec:KSlaw}

\begin{figure*}
    \centering
	\includegraphics[trim={0 30 0 5}, width=\textwidth]{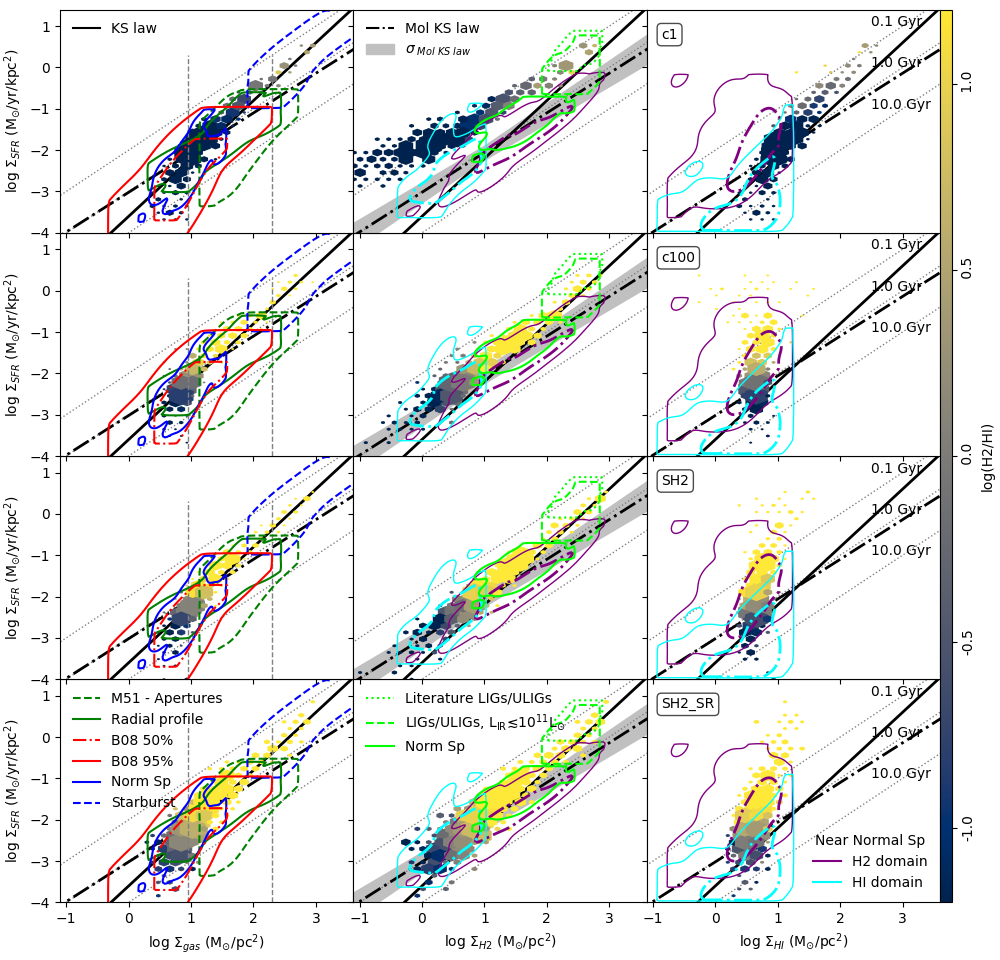}
    \caption{KS law for c1, c100, SH2 and SH2\_SR. We show the average relation for the total gas, H$_2$ and HI only, and compare them with the original (KS, solid black line) and molecular (mol-KS, black dash-dotted line) Kennicutt-Schmith law. The grey band in the central column, is a dispersion of $1\sigma$ for the mol-KS law estimated by \citet{Bigiel2008} and the red contours in the first column enclose the 50\% (dashed) and 95\% (solid) of their observational data. While the dashed grey vertical lines delimit three different regimes of SF formation defined by them.
    Also in the first column, the additional contours enclose 95\% of: measurements in individual apertures (green dashed) in M51 \citep{Kennicutt2007}, data points from radial profiles (green solid) from M51 \citep{Schuster2007}, NGC 4736, NGC 5055 \citep{Wong.Blitz2002} and NGC 6946 \citep{Crosthwaite.Turner2007}. Furthermore, we include the disc-averaged measurements from normal spiral galaxies (blue solid) and starburst galaxies (blue dashed) from \citet{Kennicutt1998}.
    In rest columns, the contours contain the 50\% (dash-dotted) and 95\% (solid) of data of \citet{Schruba2011} for H$_2$ (violet) and HI (cyan) dominated regions. And, in the middle columns, the green contours enclose the 95\% of data of \citet{Gao.Solomon2004}, for ULIGs available from literature (dotted), LIGs and ULIGs with $\mathrm{L_{IR}} \lesssim 10^{11} \mathrm{L}_{\odot}$ (dashed) and less luminous normal spiral galaxies (solid).
    In all panels, the overplotted diagonal black dotted lines correspond to 0.1, 1, and 10 Gyr constant depletion times (defined as $\tau_{\mathrm{d}} = \Sigma_{\mathrm{SFR}}/\Sigma_{\mathrm{X}}$, with X = HI+H$_2$, H$_2$ or HI), the colour scale points out the mean H$_2$-to-HI surface density ratio for each bin and the size of the cells showing the distribution of the simulated data represents the number of data within each of these.}
    \label{fig:KS_law}
\end{figure*}

In this section the KS law is used to test the performance of our models. Figure ~\ref{fig:KS_law} shows $\Sigma_{\mathrm{SFR}}$ as a function of total gas surface density ($\Sigma_{\mathrm{gas}}$), $\Sigma_{\mathrm{H_2}}$ and  $\Sigma_{\mathrm{HI}}$ for four experiments. They all assume the J75 scheme for H$_2$-dust formation. Two of them (c1 and c100) employ the standard SF scheme with $C_{\rho}=1$ and $C_{\rho}=100$, respectively, while SH2 and SH2\_SR adopt the SF scheme based on the 'self-gravity' criterion and H$_2$-dependent probability for SF scheme (with $C_{\rho}=100$). SH2\_SR also includes the stellar radiation feedback.

To compare with observations, we take the data of \citet{Schruba2011} from a sample of 33 nearby star-forming disc galaxies, with a mean metallicity of $\mathrm{12+log(O/H) \sim 8.7\ dex\ (\sigma = 0.14\ dex)}$. This is similar to the metallicity of our simulated galaxy measured at the $r_{83}$, $\mathrm{12+log(O/H) \sim 8.57}$ dex.
These authors focus their analysis on data stacked in bins of galactocentric radius, specifically in $15''$ wide tilted rings. This width varies depending on the source distance. For our comparison we adopt ${\sim} 220$ pc for its closest objects (3 Mpc). We employ a similar radial binning strategy, computing surface densities up to $1.2 r_{83}\  ({\sim} 12\ \mathrm{kpc})$ in 200 pc wide concentric rings. Then, the colour scale in Fig~\ref{fig:KS_law} indicates the mean H$_2$-to-HI surface density ratio for each bin, and the size of the cells showing the distribution of the simulated data represents the number of contributions to  each bin

We found the $\Sigma_{\mathrm{SFR}} - \Sigma_{\mathrm{gas}}$ relation to be in very good agreement with observations by \citet[][red contours in Fig~\ref{fig:KS_law}]{Bigiel2008,Bigiel2010}. These authors defined three different SF formation regimes in the $\Sigma_{\mathrm{SFR}} - \Sigma_{\mathrm{gas}}$ plane (delimited by vertical dashed lines). The first zone is dominated by HI up to a saturation threshold of 9 M$_{\odot}$ pc$^{-2}$. The second one is dominated by H$_2$ and the third one corresponds to the starburst activity. The last two regimes are delimited by a transmission band of around 200 M$_{\odot}$ pc$^{-2}$, where by increasing the density, the free-fall time is reduced within the gas cloud and can probably generate a more efficient SF activity. In addition, we include the literature measurements collected by \citet{Bigiel2008}.

In the analysed models, the simulated $\Sigma_{\mathrm{SFR}} - \Sigma_{\mathrm{gas}}$ relation reproduces the lower threshold for SF, and also the trends for normal spirals as well as starbursts galaxies, which indicates that they exhibit a strong SF activity, in some of their regions, in agreement with the trends presented by the temporal evolution of their SFRs (Fig~\ref{fig:sfr}).
The KS law obtained by using $\Sigma_{\mathrm{gas}}$ is almost unaffected by the clumping factor. However, the H$_2$-to-HI ratio varies significantly because of the impact on the molecular formation. The molecular KS law is displaced by roughly an order of magnitude between c1 and c100. This result suggests that the sub-grid clumping factor is working as we excepted, allowing us to improve the performance of our model in the very high density regions for the resolution of this IC \citep{Gnedin2009,Lupi2018}.
Furthermore, for c100, the transition zone $(\mathrm{\log(H_2/HI) \sim 0})$ is narrower than that of SH2, which is narrower than that SH2\_SR. For the last two schemes, this behaviour can be explained by the H$_2$ dependence of the SFR algorithm, and by the presence of the radiative feedback that causes the destruction of molecular hydrogen.

The molecular KS law ($\Sigma_{\mathrm{SFR}} - \Sigma_{\mathrm{H_2}}$) is also explored.
With a clumping factor $C_\rho = 100$, this correlation follows very well the observed one,  following the lines of constant depletion time (i.e. defined as the inverse of SFE) for $\tau_{\mathrm{d}} \sim 2$ Gyr  \citep{Bigiel2008, Leroy2008}, in agreement with  observations for normal spiral galaxies \citep[e.g.][]{Gao.Solomon2004, Schruba2011}. With this
$C_\rho$, the KS law is also well reproduced for  the lower density tail of $\Sigma_{\mathrm{H_2}}$. Similar behaviours are detected for SH2\_SR although these relations show slightly  larger scatter that contributes to better match the observed contours. \citet{Bigiel2008} interpret the linear relation and constant depletion times as evidence for approximately uniform SF activity in giant molecular clouds. These authors also suggest that $\Sigma_{\mathrm{H_2}}$ may be a better measure of the filling fraction of GMC than of the changing conditions in the molecular gas.

For high density regions, our models overestimate the $\Sigma_{\mathrm{SFR}}$ with respect to a constant $\tau_{\mathrm{d}}$, overlapping with the starburst regime. In fact, as can be seen in Fig~\ref{fig:KS_law}, the agreement with observational data of luminous and ultra-luminous infrared galaxies (LIGs/ULIGs) \citep{Gao.Solomon2004} is remarkably good.

The difference between SH2 and SH2\_SR models is mainly limited to larger dispersion in the distributions of the latter, as exhibit their surface density distributions represented by the cell sizes in Fig~\ref{fig:KS_law}.
This reflects the dissociating effect of stellar radiation, although it has a weak impact on the global trend. Nevertheless, due to the similarity between both models, further studies are required to understand whether the correlation between SF and H$_2$ content implies causality.

\citet{Byrne2019} reported similar results for their SF schemes, since all of them are consistent with the KS law with only slight differences with the observable properties of the resulting galaxies. However, these authors claim that  their effects would be stronger in simulations focused on the very early universe or capable of resolving the internal structure of star-forming clouds.

Finally, the $\Sigma_{\mathrm{SFR}} - \Sigma_{\mathrm{HI}}$ relation agree with observational results reported by \citet{Bigiel2008,Bigiel2010} and \citet{Schruba2011}.
Our initial condition for the disc component, including the stellar and gas phase, follows an exponential profile. 
Since simulated galaxies evolve in isolation without experiencing any external disturbance, we can assume that the initial exponential profiles imprinted in the stellar and gas components does not undergo significant changes, so that the internal regions are on average denser than the external ones and therefore they satisfy the conditions to form stars sooner and more efficiently.
Indeed, the higher level of SF activity corresponds to the central regions in these experiments.
\citet{Bigiel2008,Bigiel2010} found that SF in outer disc regions is very inefficient
compared to SF inside the optical discs. In addition they observed HI depletion times of 100 Gyr for most of their data, and regions with local $\tau_{\mathrm{d}}<10$ Gyr are rare to find. 
The disagreement with our simulations at low gas densities $(\lesssim 3\ \mathrm{M_{\odot}\ pc^{-2}})$ seems to indicate that further adjustment of our chemical model and/or the SF scheme is required to improve the modelling at very low gas densities.

\citet{Bigiel2010} conclude that, on average, outer disc contribute only about $10\%$ to the total SFR of a galaxy and the lack of importance of in situ SF means that the massive extended gas distributions observed in many nearby galaxies can be interpreted as a potential fuel source for inner disc SF. The short depletion times in these regions ($\tau_{\mathrm{d}} \sim 2$ Gyr) imply that this source is required as well as the presence of mechanisms such as radial gas flows that replenish the innermost areas.

\subsection{Dependence on metallicity} \label{sec:metallicity}

\begin{figure*}
    \centering
	\includegraphics[trim={0 30 0 0}, width=\textwidth]{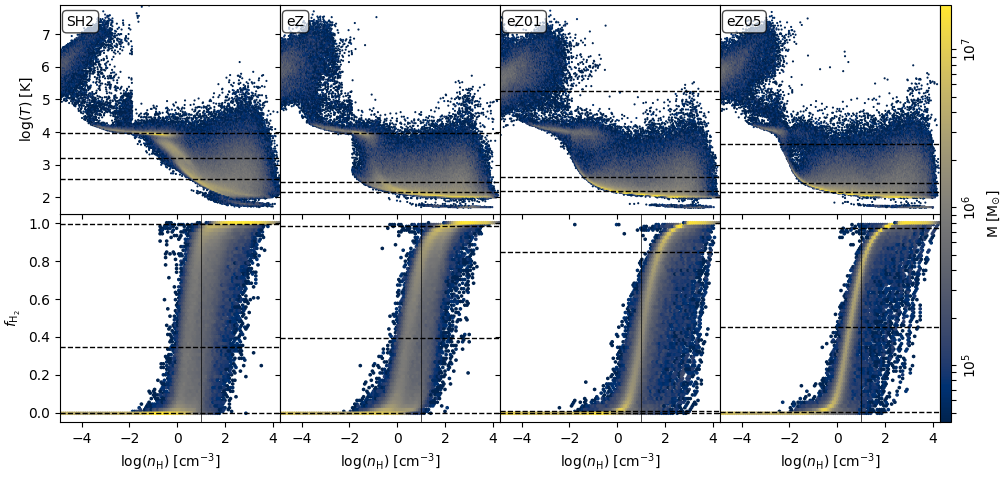}
    \caption{Temperature and H$_2$ mass fraction as a function of  $n_{\mathrm{H}}$ (cm$^{-3}$), for  SH2, eZ, eZ01 and eZ05 (left to right, respectively).
    The dashed lines indicate the 25, 50 and 75 percentiles for the distributions of temperature (upper panels) and H$_2$ mass fraction (lower panels).
    The solid vertical line in lower panels denotes $n\mathrm{_{H} = 10\ cm^{-3}}$.}
    \label{fig:dens_temp_fH2}
\end{figure*}

\begin{figure}
    \centering
	\includegraphics[trim={0 5 0 10},width=.47\textwidth]{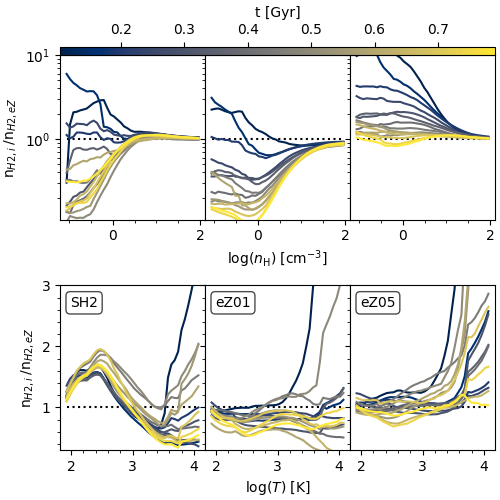}
    \caption{Temporal evolution of the H$_2$ abundances ratio $(n_{\mathrm{H_2,i}}/n_{\mathrm{H_2,eZ}})$ of i = SH2, eZ01 and eZ05 models with respect to eZ as function of $n_{\mathrm{H}}$ (upper panel) and the temperature (lower panel). The dotted line indicates when the H$_2$ abundance of each model is equal to that of eZ $(n_{\mathrm{H_2,i}}/n_{\mathrm{H_2,eZ}} = 1)$.}
    \label{fig:metal_H2}
\end{figure}

In this section, we analyse the dependence of H$_2$ formation on gas metallicity. Our goal is to understand the effects of considering a fixed metallicity for the gas or a self-consistent chemical enrichment by SNae that produce the variation of the gas metallicity over time. All the experiments discussed in this section adopt the self-gravitation model with H$_2$ dependence (see Table 1).

We performed two different experiments adopting a uniform distribution of chemical abundances per gas particle, consistent with a metallicity of 0.1 $Z_{\odot}$ (eZ01) and 0.5  $Z_{\odot}$ (eZ05). However, this condition obstructs us from implementing non-equilibrium metal cooling schemes for $T<10^4$ K, because fixing the metallicity prevents the evolution of the numerical densities of the metal species relevant for cooling as required by our scheme. To solve this, we extended the equilibrium metal cooling range to $T < 10^4$ K, following  \citet{Lupi2018}.  This scheme was implemented for these two experiments and a third run (eZ) that uses the full chemical evolution model. eZ  includes SNII and SNIa enrichment and our initial ad hoc metallicity profile (see sec~\ref{sec:ic}). The later run is used  as a benchmark to assess the effects of assuming the fixed abundance under similar conditions for the cooling process.

There are several physical phenomena linked to the level of chemical enrichment in the gas phase. Metal cooling increases with metallicity because of the enhanced free electron density. In our equilibrium scheme it scales linearly with $Z$ \citep{Shen2010,Shen2013}.
High metallicities also imply high dust-to-gas ratios, which in this work are assumed to scale as $D = D (Z/Z_{\odot})$ (see sec~\ref{sec:model}). Consequently, we also expect a larger abundance of molecular hydrogen, and stronger molecular-hydrogen cooling (which, however, at the densities and metallicities considered in this work, is subdominant with respect to metal cooling). 
In agreement with the above expectations, the gas density-temperature diagrams (top panel of Fig.~\ref{fig:dens_temp_fH2}) show how the cold gas fraction increases with increasing metallicity (the dashed horizontal lines depict the gas temperature reached by the  25, 50, 75 per cent of the gas mass). 

In particular, both  eZ01 and eZ05 show the denser 25\%  of total gas mass  located below ${\sim} 3 \times 10^2$ K, in the  H$_2$-dominated region, whose contribution to the gas cooling is relevant.
The denser 50\% of cold gas mass also spans similar ranges of density in both simulations but in eZ05 it is colder than in eZ01, where this gas is below ${\sim} 4 \times 10^2$ K and ${\sim} 7 \times 10^2$ K, respectively. 
Finally, the temperatures of gas mass distributions from  50 to 75 percentiles show larger differences, since in eZ05 they do not exceed the ${\sim} 5 \times 10^3$ K while in eZ01, achieve ${\sim} 1 \times 10^5$ K.
In the lowest metallicity model (eZ01), there is a weaker impact of metal cooling and also of H$_2$ cooling due to the lower H$_2$ mass formed. However, comparing eZ with eZ01 and eZ05 below $10^3$ K and for high density regions, we note that the direct dependence of the cooling processes on metals does not appear to be very strong.
As we can seen in the lower panels of Fig.~\ref{fig:dens_temp_fH2}, the impact of metallicity on the H$_2$ abundance at a given $n_{\mathrm{H}}$ is very clear for the higher metallicity run.

As expected from the direct link between metallicity, dust, and H$_2$ formation, the range of densities at which the gas starts transitioning from atomic to molecular (i.e. the transition zone) slightly shifts towards lower densities with increasing metallicity and becomes more concentrated \citep{Christensen2012}. In our experiments these changes are small since the difference in metallicity between both simulations is only a factor of five. Nevertheless, the amount of gas with high $f_{\mathrm{H_2}}$ increases.
For example, ${\sim} 25$ per cent of the gas mass has a $f_{\mathrm{H_2}} \gtrapprox 0.90$ for eZ01 and $f_{\mathrm{H_2}} \gtrapprox 0.96$ for eZ05, while 50  per cent of gas mas  has $f_{\mathrm{H_2}} \gtrapprox 0.10$ and $f_{\mathrm{H_2}} \gtrapprox 0.50$ for each simulation, respectively (see dashed lines in lower panels of Fig.~\ref{fig:dens_temp_fH2}). We also note a wide spread of densities in the region of gas transitioning from atomic to molecular.

If we focus on SH2 and eZ (first two columns in Fig.~\ref{fig:dens_temp_fH2}), we find that a metal cooling process assuming thermochemical equilibrium substantially affects the $n_{\mathrm{H}}$ and temperature distributions mainly at low densities, e.i. $n_{\mathrm{H}} \lesssim 10\ \mathrm{cm}^{-3}$. We recall that this $n_{\mathrm{H}}$ threshold denotes the mean density of the transition zone between HI and H$_2$. However, for the higher $n_{\mathrm{H}}$ there are not large changes in the H$_2$ fractions. This can be explained because the main formation channel for H$_2$ is the catalysis on the surface of the dust which depends on the metallicity and $n_{\mathrm{H}}$.

In addition, comparing the temporal evolution of SH2 (which has an initial metallicity gradient corresponding to an average gradient observed for z = 2) with a simulation that starts with primordial gas (LR, see Table~\ref{tab:runs_table}), we observe that the latter requires ${\sim} 0.45$ Gyr to reach a $\Sigma_{\mathrm{H_2}}$ similar to that of SH2 in $r < 0.5 r_{83}$, while for $r > r_{83}$ displays a $\Sigma_{\mathrm{H_2}}$ always smaller (for a factor of ${\sim} 5$) for at least 2 Gyr.

A more quantitative analysis is shown in Fig~\ref{fig:metal_H2}. Here, for each model (SH2, eZ01 and eZ05), we calculate the distribution of H$_2$ number density normalised by the corresponding value in eZ ($n_{\mathrm{H_2,i}}/n_{\mathrm{H_2,eZ}}$; i denotes the different models) as a function of the temperature and the $n_{\mathrm{H}}$ for different evolution times. We compare these distributions, evaluating the ratio of H$_2$ abundance and the time variation. We remark that, in addition to employing thermochemical equilibrium cooling rates, eZ includes the full chemical evolution driven by SN feedback so that the metallicity of the ISM will increase with time.

First we compare the evolution of the $n_{\mathrm{H_2}}$ between SH2 and eZ which allow us to evaluate the impact of non-equilibrium cooling. As it can be seen from Fig~\ref{fig:metal_H2} (left panels) as a function of $n_{\mathrm{HI}}$ there are no significant differences for $n_{\mathrm{HI}}>10$ cm$^{-3}$. However for lower densities, eZ underestimates the H$_2$ production in the initial times, but as the enrichment increases with the temporal evolution of the system, eZ begins to overestimate the H$_2$ content.
For $T \lesssim 1000$ K, eZ underestimates $n_{\mathrm{H_2}}$ while overestimating it for higher temperatures (except for a short short period of ${\sim} 0.1$ Gyr, from 0.4 to 0.5 Gyr).

For eZ01 and eZ05, we observe that the production of H$_2$ in very dense and cold regions is similar in all cases. This indicates that these regions are prompted to molecular H formation regardless of metal abundance. However, as \citet{Capelo2018} also observed, if we consider lower density regions, the metallicity begins to play an important role due to its influence on the schemes chosen here to model the H$_2$ formation in dust. Thus, as the metallicity increases in eZ (due to SN feedback), its H$_2$ production evolves approaching that of eZ05 and moving away from eZ01.
As a function of the temperature,  we find again that the H$_2$ abundance rises with increasing metal content and exhibits a large dispersion as shown in Fig.~\ref{fig:metal_H2}. However, the larger impact of metallicity on the synthesis of H$_2$ is detected in regions with temperatures $T \gtrsim 1 \times 10^3$ K, while  in very cold and dense regions ($T \lesssim 1 \times 10^3 $K) the effects are less important.  
This indicates that for the H$_2$ formation in the aforementioned cold and dense areas, the dominant process is the H$_2$ catalysis on the dust grains surface. As the models used for this mechanism assume  a direct proportionality with $Z/Z_{\odot}$ and $n_{\mathrm{H}}$, then at high density regions $(n_{\mathrm{H}} \gtrsim 10)$ the last quantity dominates widely over the metal abundances $(Z_{\odot} \lesssim 1)$ (see sec~\ref{sec:model}).

Finally, we consider the impact of metallicity and H$_2$ formation on the morphology and spatial distribution of H$_2$ in gas-phase disc (not shown here, see Fig A in supplementary online material).
The models with fixed metallicity all over the discs (eZ01 and eZ05) show a more widespread molecular formation. This is related to the temperature pattern shown in Fig~\ref{fig:metal_H2} in which low temperatures extend more evenly and to the outermost disc regions for higher metallicities.
We can still individualise stellar and molecular formation in clumps. However, compared to the SH2 model (Fig.~\ref{fig:sf_colormap}), these are smaller and more tenuous, and are found in large numbers and more evenly distributed across the disc.

\subsection{Considerations on numerical resolution} \label{sec:num_resol}

\begin{figure}
    \centering
	\includegraphics[trim={0 0 0 25}, width=.47\textwidth]{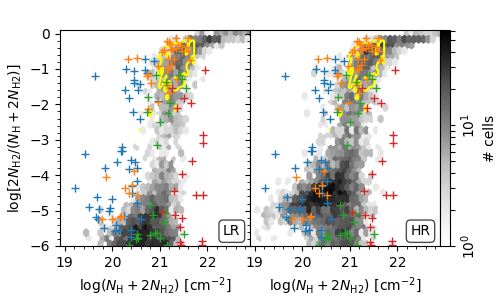}
	\includegraphics[trim={0 0 0 0}, width=.47\textwidth]{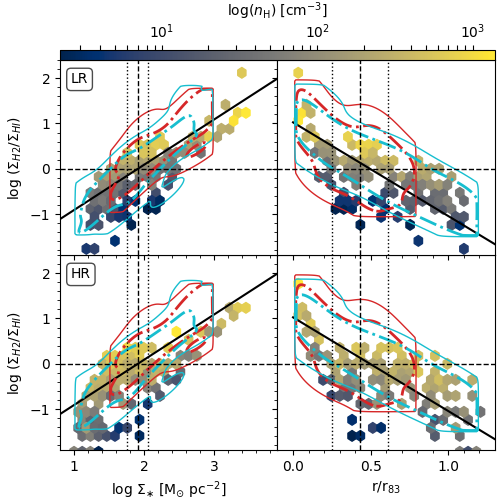}
    \caption{Observational comparisons for runs LR and HR. Top panels: H$_2$ column density fraction (Idem Fig.~\ref{fig:SH2_obs1}). Bottom panels: H$_2$-to-HI ratio, $R_{\mathrm{mol}} = \Sigma_{\mathrm{H_2}} / \Sigma_{\mathrm{HI}}$, as a function of $\Sigma_{\ast}$ (left) and radius (right - Idem Fig.~\ref{fig:SH2_obs2}).}
    \label{fig:HR_obs}
\end{figure}

\begin{figure*}
    \centering
	\includegraphics[trim={0 20 0 0}, width=\textwidth]{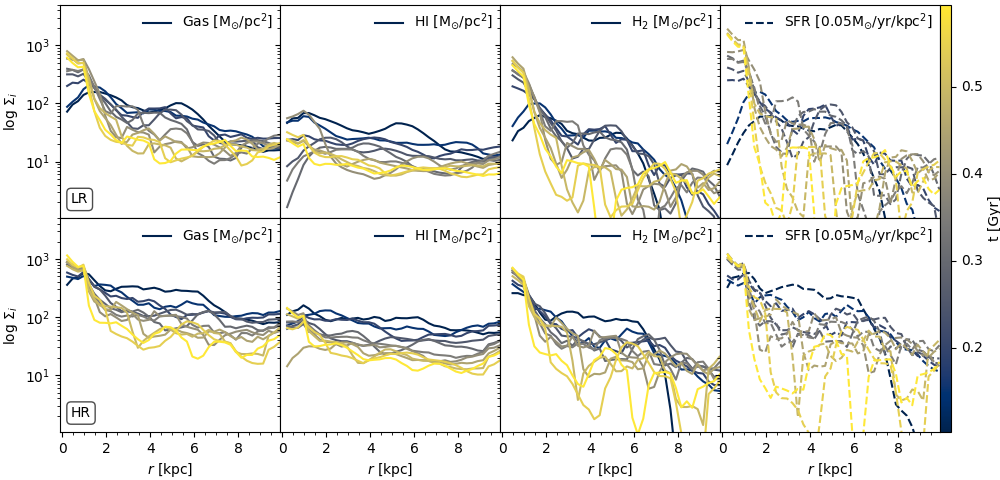}
    \caption{Surface density profiles of total gas, stars, HI, H$_2$ and SFR for LR (upper panels) and HR (lower panels) experiments.}
    \label{fig:LR_vs_HR_prof}
\end{figure*}

\begin{figure*}
    \centering
	\includegraphics[trim={0 20 0 0}, width=\textwidth]{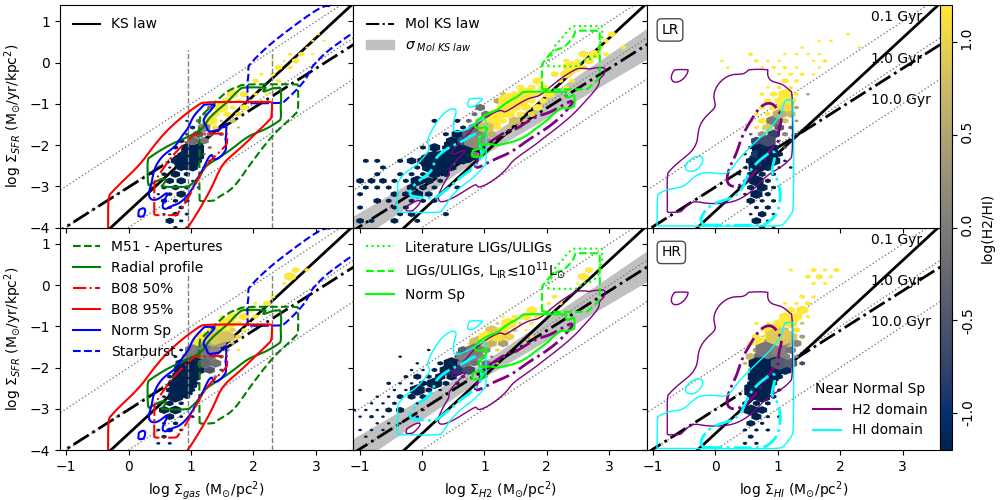}
    \caption{KS laws for runs LR (upper panels) and HR (lower panels). We show the average relation for the total gas, H$_2$ and HI only, and compare them with the original (KS, solid black line) and molecular (mol-KS, black dash-dotted line) Kennicutt-Schmith law. This figure is equivalent to Fig.~\ref{fig:KS_law} for a different set of models.}
    \label{fig:KS_law_HR}
\end{figure*}

In order to analyse the impact of numerical resolution, we present the results of two runs with different resolution, hereafter high (HR) and low (LR) resolution. The ICs for this analysis are described at the end of section~\ref{sec:ic}. Simulations start from primordial gas without adopting an initial metallicity gradient and these simulations used our best model: SH2 and J75 schemes, with $C_{\rho} = 10$ and 100 for the HR and LR respectively (see table~\ref{tab:runs_table}).
Recall that the clumping factor is introduced to take into account the effects of limited numerical resolution. Hence, for the HR run a lower value, $C_{\rho} = 10$, is required in order to reproduce observational trends (Fig.~\ref{fig:HR_obs}). 

For simplicity, we do not include the complete analysis  carried out for low resolution runs, but these studies have been fully done on the HR run. Instead, we limit the discussion to examine and contrast the radial distributions of total gas, HI, H$_2$ and the SF activity present in LR and HR (Fig.~\ref{fig:LR_vs_HR_prof}), and the behaviour of the KS laws (Fig.~\ref{fig:KS_law_HR}). Our goal is not to reproduce any specific galaxy, but to evaluate the impact of numerical resolution using our reference model to ensure that our trends are robust.

We compare the radial profiles of H$_2$ and SFR estimated for our simulations with the observational ones by \citet{Leroy2008} and \citet{Gallagher2018}. The general trend in these observations is that the both profiles reach the maximum densities in the galactic centre and decay to the outskirts of the disc, while HI is approximately constant across the disc.
Our results are in very good agreement with the observations and since the clumping factor differs by an order of magnitude between the two runs, it is clear that an increase in resolution affects the model of H$_2$ formation on dust, since  higher density regions can be better resolved as well as the chemical processes that take place in them. However, the tuning of this parameter allows us to recover a good match with observations.

Additionally, the H$_2$ and SFR profiles and their evolution are very similar in the two simulations, on average.
By looking in detail, we note that initially, the surface density of H$_2$ exceeds that of HI, which increases slightly as the system evolves. In addition, the SFR of the LR run reaches marginally higher values in the central region than HR, and falls faster in the intermediate disc regions. Probably, these higher central SFRs, on average, are due to the limitations imposed by the resolution, which establishes the mass of each gas particle that will form new stars, obtaining a formation in discrete quantities whose stellar mass is greater for lower resolutions. As the numerical resolution improves, this effect weakens and the simulation better reproduces  observed distributions of the H$_2$ column density fractions and the H$_2$-to-HI ratios as a function of $\Sigma_{\ast}$ and radius (see Fig~\ref{fig:HR_obs}). However, in both simulations, the H$_2$ profile reaches its peak in the galactic centre and falls towards the outer disc, while the HI profile remains relatively flat throughout the disc. The SFR also peaks in the centre and trends in its profile largely follows those of H$_2$.
Finally the stellar distributions are very similar for both resolution runs.

Hence, these trends show that, although the distributions of chemical species (HI and H$_2$) and star-forming regions are slightly smoother with increasing resolution,  both runs provides similar distributions which are globally consistent with observations.

\section{Summary} \label{sec:4}

In this work, we developed a new implementation of {\small P-GADGET3} that includes the chemistry package {\small KROME}. {\small P-GADGET3-K} allows a self-consistent description of H$_2$ formation as baryons cool and form new stars while the ISM is chemically enriched by SNIa and SNII. 
Our code considers energy feedback by SNae, an extragalactic UV  and local stellar radiation fields.
We run a set of isolated galaxies to evaluate the performance of {\small P-GADGET3-K} and the relevance of different physical prescriptions and their free parameters (H$_2$ formation via dust grains, clumping factor and metallicity ) as well as the impact of H$_2$ on SF activity.
Special care was given to the initial gas metallicity so that the gas component starts with a metallicity profile in agreement with observations at $z {\sim} 2$, considering that chemical abundances affects both the cooling rates and H$_2$ formation. We explore two possible models for the formation of H$_2$ via dust grains, \citet{Jura1975} and \citet{Tomassetti2014}.

Our results can be summarised as follows:

\begin{itemize} 
\item We find that the characteristics of simulated systems are in very good agreement with a variety of observable properties of molecular gas (Fig.~\ref{fig:leroy_2008}). $R_{\mathrm{mol}}$ covers a similar range of values as observations and behaves as a continuous function spanning from the H$_2$-dominated (${\sim} 10$) to HI-dominated (${\sim} 0.1$) regions of the ISM. Particularly, for the adopted numerical resolution, the runs with $C_{\rho} = 100$ show the best fits to observational distributions as a function of  $\Sigma_{\mathrm SFR}$ and effective radius (Fig.~\ref{fig:leroy_2008}). 

The H$_2$ fractions  is comparable to the observations on average (Fig.~\ref{fig:obs_2}), although the T14 dust model is found to produce less H$_2$ at a given $N_{\mathrm{H}}$. As expected the profiles of $\Sigma_{\mathrm H_2}$, $\Sigma_{\mathrm HI}$ and $\Sigma_{\mathrm SFR}$ vary with the clumping factor and numerical resolution. The ISM in the standard experiments with low $({<} 100)$ or variable (cv) clumping factor are more dominated by HI and they do not fit these observations well.
On the other hand, the observed range of column densities are better reproduced by the experiments with a variable $C_{\rho}$ and J75 or a $C_{\rho}=100$ and T14, that is, where the H$_2$-dust schemes depend on some local thermodynamic properties of the gas.
Therefore, in a future work, we will analyse in detail  these schemes, seeking to understand the relationship between the clumping factor, the thermodynamical properties of the gas and the numerical resolution, to obtain guidelines for choosing the most appropriate model.

\item In our model with local stellar radiation field, SH2\_SR, the simulated distribution of $N_{\mathrm{H_2}}$ with respect to $N_{\mathrm{H}} + 2N_{\mathrm{H_2}}$ improves and reproduce values within the entire observational range set by MW, LMC and SMC.
However, the $\Sigma_{\mathrm{H_2}}$ distributions or KS relations are not significantly different when compared to SH2, since the greatest impact of radiation occurs in less dense regions where the effects of attenuation and self-shielding are not dominant where star formation is less active.

\item We find a very good agreement between the abundances and distributions of HI, H$_2$, SFR and stellar mass of the LR and HR experiments (Fig.~\ref{fig:HR_obs}, \ref{fig:LR_vs_HR_prof}). The effects of low resolution can be seen in the central regions where LR exhibits higher levels of activity. As expected, the HR required a lower clumping factor to reproduce observations.

\item The density-temperature diagrams show that simulated galaxies follow the same trends, on average, but the extreme values that the temperature and/or the density can reach depend on the chemical content and numerical resolution. This, together with the $C_{\rho}$, contribute to shape the phase diagrams for H$_2$.

\item We tested two different SF schemes and explore the gas behaviour and, the molecular and HI KS laws. For the standard SF model, based on a density and temperature thresholds, we run experiments varying the clumping factor and the H$_2$-dust model. Increasing the clumping factor generates an increase of the H$_2$ abundance as self-shielding is more efficient. This produces a molecular KS law with a slope slightly closer to the observed one, while the non-use of a clumping factor increases the differences between observations and simulations. We also found that the use of the J75 scheme results in a larger fraction of H$_2$ than the T14 model, with the same clumping factor. The SFR history produced by the standard SF scheme shows fluctuating behaviour with many significant peaks and valleys interspersed. Nevertheless, the second model, based on the "auto-gravity" criterion \citep{Hopkins2013b, Hopkins2015} and a H$_2$-dependent probability of SF, shows a SF activity smoother over time but with a similar global trend to the previous scheme.
We do not find large differences in the SFR and global properties of ISM when using a H$_2$-dependent SF algorithm. However, the SH2 and SH2\_SR runs reproduce observational trends slightly better (Fig.~\ref{fig:SH2_obs1} and ~\ref{fig:SH2_obs2}). This is encouraging given the simplicity of the algorithm and will be further analysed in a future work.

\item On the other hand, the modelling of the chemical evolution of metallic species (CI, CII, OI, OII, SiI, SiII and SiIII) and non-equilibrium cooling has a significant effect on gas thermodynamics, chemical abundances and morphology of the galaxy, in agreement with previous works.
Compare to a model that assumes thermochemical equilibrium, the non-equilibrium scheme produces a wider distribution of the gas with temperatures around ${\sim} 10^3 K$. It also generates  higher fractions of molecular gas at low temperatures. The spatial distribution of the molecular components is more extended in the non-equilibrium scheme. However, the differences in gas thermodynamics are typically small for the cold and dense regions where stars are formed,  implying similar levels of  SF activities.

\item As expected, the metallicity of gas has direct consequences on its cooling rates and the H$_2$ formation. A model with higher initial metallicity produces a colder gas and a larger fraction of molecular gas than one with primordial or low metallicity. However the metal enrichment by SNae tend to dilute the initial metallic discrepancy between runs. 
We find that the presence of a pre-set metallicity gradient allows a better description of the H$_2$-to-HI ratio and avoids the low-density tail in the molecular KS relation.
\end{itemize}

In summary, our updated code {\small P-GADGET3-K} is a powerful numerical tool that will allow us to address, in a more detailed and physical way, the regulation of SF and the gas cycle during galactic evolution.

\section*{Acknowledgements}
The authors would like to thank the anonymous reviewers for their insightful comments.
This work used the RAGNAR cluster, funded by Fondecyt 1150334 and Universidad Andres Bello, and the Geryon2 cluster at the Centro de Astro-Ingenier\'ia UC. 
PBT acknowledges partial supports from Fondecyt 120073/2020 and the Galnet Network supported by CONICYT. 
SB and DRGS thanks for funding via Conicyt PIA ACT172033. 
SB ackonwledges support from Fondecyt Iniciaci\'on 11170268 and
and DRGS thanks for funding via Conicyt Programa de Astronom\'a Fondo Quimal 2017 QUIMAL170001.
DRGS and SB acknowledge financial support from the Millenium Nucleus NCN19\_058 (TITANs).

\section*{Data Availability}
The data underlying this article will be shared on reasonable request
to the corresponding author.


\bibliographystyle{mnras}
\bibliography{References}

\begin{thebibliography}{}
\makeatletter
\relax
\def\mn@urlcharsother{\let\do\@makeother \do\$\do\&\do\#\do\^\do\_\do\%\do\~}
\def\mn@doi{\begingroup\mn@urlcharsother \@ifnextchar [ {\mn@doi@}
  {\mn@doi@[]}}
\def\mn@doi@[#1]#2{\def\@tempa{#1}\ifx\@tempa\@empty \href
  {http://dx.doi.org/#2} {doi:#2}\else \href {http://dx.doi.org/#2} {#1}\fi
  \endgroup}
\def\mn@eprint#1#2{\mn@eprint@#1:#2::\@nil}
\def\mn@eprint@arXiv#1{\href {http://arxiv.org/abs/#1} {{\tt arXiv:#1}}}
\def\mn@eprint@dblp#1{\href {http://dblp.uni-trier.de/rec/bibtex/#1.xml}
  {dblp:#1}}
\def\mn@eprint@#1:#2:#3:#4\@nil{\def\@tempa {#1}\def\@tempb {#2}\def\@tempc
  {#3}\ifx \@tempc \@empty \let \@tempc \@tempb \let \@tempb \@tempa \fi \ifx
  \@tempb \@empty \def\@tempb {arXiv}\fi \@ifundefined
  {mn@eprint@\@tempb}{\@tempb:\@tempc}{\expandafter \expandafter \csname
  mn@eprint@\@tempb\endcsname \expandafter{\@tempc}}}

\bibitem[\protect\citeauthoryear{ALMA-Partnership et~al.,}{ALMA-Partnership
  et~al.}{2015}]{ALMAPartnership2015}
ALMA-Partnership et~al., 2015, \mn@doi [ApJ] {10.1088/2041-8205/808/1/L1}, 808,
  L1

\bibitem[\protect\citeauthoryear{Abel, Anninos, Zhang  \& Norman}{Abel
  et~al.}{1997}]{Abel1997}
Abel T.,  Anninos P.,  Zhang Y.,   Norman M.~L.,  1997, \mn@doi [New Astronomy]
  {10.1016/S1384-1076(97)00010-9}, 2, 181

\bibitem[\protect\citeauthoryear{{Aoyama} et~al.,}{{Aoyama}
  et~al.}{2017}]{Aoyama2017}
{Aoyama} S.,  et~al., 2017, \mn@doi [\mnras] {10.1093/mnras/stw3061}, \href
  {https://ui.adsabs.harvard.edu/abs/2017MNRAS.466..105A} {466, 105}

\bibitem[\protect\citeauthoryear{Arth et~al.,}{Arth et~al.}{2019}]{Arth2019}
Arth A.,  et~al., 2019, arXiv:1907.11250 [astro-ph]

\bibitem[\protect\citeauthoryear{Bakes \& Tielens}{Bakes \&
  Tielens}{1994}]{Bakes.Tielens1994}
Bakes E. L.~O.,  Tielens A. G. G.~M.,  1994, ApJ, 427, 822

\bibitem[\protect\citeauthoryear{Balsara}{Balsara}{1995}]{Balsara1995}
Balsara D.~S.,  1995, \mn@doi [J. Comput. Phys.]
  {10.1016/S0021-9991(95)90221-X}, 121, 357

\bibitem[\protect\citeauthoryear{Beck et~al.,}{Beck et~al.}{2016}]{Beck2016}
Beck A.~M.,  et~al., 2016, \mn@doi [MNRAS] {10.1093/mnras/stv2443}, 455, 2110

\bibitem[\protect\citeauthoryear{Bigiel et~al.,}{Bigiel
  et~al.}{2008}]{Bigiel2008}
Bigiel F.,  et~al., 2008, \mn@doi [AJ] {10.1088/0004-6256/136/6/2846}, 136,
  2846

\bibitem[\protect\citeauthoryear{Bigiel et~al.,}{Bigiel
  et~al.}{2010}]{Bigiel2010}
Bigiel F.,  et~al., 2010, \mn@doi [AJ] {10.1088/0004-6256/140/5/1194}, 140,
  1194

\bibitem[\protect\citeauthoryear{Blitz \& Rosolowsky}{Blitz \&
  Rosolowsky}{2006}]{Blitz.Rosolowsky2006}
Blitz L.,  Rosolowsky E.,  2006, \mn@doi [ApJ] {10.1086/505417}, 650, 933

\bibitem[\protect\citeauthoryear{Bolatto, Wolfire  \& Leroy}{Bolatto
  et~al.}{2013}]{Bolatto2013}
Bolatto A.~D.,  Wolfire M.,   Leroy A.~K.,  2013, \mn@doi [Annu. Rev. Astron.
  Astrophys.] {10.1146/annurev-astro-082812-140944}, 51, 207

\bibitem[\protect\citeauthoryear{{Bovino}, {Grassi}, {Schleicher}  \&
  {Latif}}{{Bovino} et~al.}{2014}]{Bovino2014}
{Bovino} S.,  {Grassi} T.,  {Schleicher} D.~R.~G.,   {Latif} M.~A.,  2014,
  \mn@doi [\apjl] {10.1088/2041-8205/790/2/L35}, \href
  {https://ui.adsabs.harvard.edu/abs/2014ApJ...790L..35B} {790, L35}

\bibitem[\protect\citeauthoryear{Bovino, Grassi, Capelo, Schleicher  \&
  Banerjee}{Bovino et~al.}{2016}]{Bovino2016}
Bovino S.,  Grassi T.,  Capelo P.~R.,  Schleicher D. R.~G.,   Banerjee R.,
  2016, \mn@doi [A\&A] {10.1051/0004-6361/201628158}, 590, A15

\bibitem[\protect\citeauthoryear{{Bromm} \& {Loeb}}{{Bromm} \&
  {Loeb}}{2003}]{Bromm.Loeb2003}
{Bromm} V.,  {Loeb} A.,  2003, \mn@doi [\nat] {10.1038/nature02071}, \href
  {https://ui.adsabs.harvard.edu/abs/2003Natur.425..812B} {425, 812}

\bibitem[\protect\citeauthoryear{{Bruzual} \& {Charlot}}{{Bruzual} \&
  {Charlot}}{2003}]{Bruzual.Charlot2003}
{Bruzual} G.,  {Charlot} S.,  2003, \mn@doi [\mnras]
  {10.1046/j.1365-8711.2003.06897.x}, \href
  {https://ui.adsabs.harvard.edu/abs/2003MNRAS.344.1000B} {344, 1000}

\bibitem[\protect\citeauthoryear{{Byrne}, {Christensen}, {Tsekitsidis},
  {Brooks}  \& {Quinn}}{{Byrne} et~al.}{2019}]{Byrne2019}
{Byrne} L.,  {Christensen} C.,  {Tsekitsidis} M.,  {Brooks} A.,   {Quinn} T.,
  2019, \mn@doi [\apj] {10.3847/1538-4357/aaf9aa}, \href
  {https://ui.adsabs.harvard.edu/abs/2019ApJ...871..213B} {871, 213}

\bibitem[\protect\citeauthoryear{Capelo, Bovino, Lupi, Schleicher  \&
  Grassi}{Capelo et~al.}{2018}]{Capelo2018}
Capelo P.~R.,  Bovino S.,  Lupi A.,  Schleicher D. R.~G.,   Grassi T.,  2018,
  \mn@doi [MNRAS] {10.1093/mnras/stx3355}, 475, 3283

\bibitem[\protect\citeauthoryear{Catinella et~al.,}{Catinella
  et~al.}{2018}]{Catinella2018}
Catinella B.,  et~al., 2018, \mn@doi [MNRAS] {10.1093/mnras/sty089}, 476, 875

\bibitem[\protect\citeauthoryear{{Cazaux} \& {Spaans}}{{Cazaux} \&
  {Spaans}}{2004}]{Cazaux.Spaans2004}
{Cazaux} S.,  {Spaans} M.,  2004, \mn@doi [\apj] {10.1086/422087}, \href
  {https://ui.adsabs.harvard.edu/abs/2004ApJ...611...40C} {611, 40}

\bibitem[\protect\citeauthoryear{{Cazaux} \& {Spaans}}{{Cazaux} \&
  {Spaans}}{2009}]{Cazaux.Spaans2009}
{Cazaux} S.,  {Spaans} M.,  2009, \mn@doi [\aap] {10.1051/0004-6361:200811302},
  \href {https://ui.adsabs.harvard.edu/abs/2009A&A...496..365C} {496, 365}

\bibitem[\protect\citeauthoryear{{Cen}}{{Cen}}{1992}]{Cen1992}
{Cen} R.,  1992, \mn@doi [\apjs] {10.1086/191630}, \href
  {https://ui.adsabs.harvard.edu/abs/1992ApJS...78..341C} {78, 341}

\bibitem[\protect\citeauthoryear{{Chen} et~al.,}{{Chen}
  et~al.}{2015}]{Chen2015}
{Chen} Y.,  et~al., 2015, \mn@doi [\mnras] {10.1093/mnras/stv1281}, \href
  {https://ui.adsabs.harvard.edu/abs/2015MNRAS.452.1068C} {452, 1068}

\bibitem[\protect\citeauthoryear{Christensen et~al.,}{Christensen
  et~al.}{2012}]{Christensen2012}
Christensen C.,  et~al., 2012, \mn@doi [MNRAS]
  {10.1111/j.1365-2966.2012.21628.x}, 425, 3058

\bibitem[\protect\citeauthoryear{Crosthwaite}{Crosthwaite}{2007}]{Crosthwaite.Turner2007}
Crosthwaite L.~P.,  2007, ApJ, 134, 16

\bibitem[\protect\citeauthoryear{Cullen \& Dehnen}{Cullen \&
  Dehnen}{2010}]{Cullen.Dehnen2010}
Cullen L.,  Dehnen W.,  2010, \mn@doi [MNRAS]
  {10.1111/j.1365-2966.2010.17158.x}, 408, 669

\bibitem[\protect\citeauthoryear{Dame, Hartmann  \& Thaddeus}{Dame
  et~al.}{2001}]{Dame2001}
Dame T.,  Hartmann D.~H.,   Thaddeus P.,  2001, ApJ, 547, 792

\bibitem[\protect\citeauthoryear{Dehnen \& Aly}{Dehnen \&
  Aly}{2012}]{Dehnen.Aly2012}
Dehnen W.,  Aly H.,  2012, \mn@doi [MNRAS] {10.1111/j.1365-2966.2012.21439.x},
  425, 1068

\bibitem[\protect\citeauthoryear{{Draine}}{{Draine}}{2003}]{Draine2003}
{Draine} B.~T.,  2003, \mn@doi [\araa]
  {10.1146/annurev.astro.41.011802.094840}, \href
  {https://ui.adsabs.harvard.edu/abs/2003ARA&A..41..241D} {41, 241}

\bibitem[\protect\citeauthoryear{Federrath, {Roman-Duval}, Klessen, Schmidt  \&
  Mac~Low}{Federrath et~al.}{2010}]{Federrath2010b}
Federrath C.,  {Roman-Duval} J.,  Klessen R.~S.,  Schmidt W.,   Mac~Low M.-M.,
  2010, \mn@doi [A\&A] {10.1051/0004-6361/200912437}, 512, A81

\bibitem[\protect\citeauthoryear{Ferland et~al.,}{Ferland
  et~al.}{1998}]{Ferland1998}
Ferland G.~J.,  et~al., 1998, \mn@doi [Publ. Astron. Soc. Pac.]
  {10.1086/316190}, 110, 761

\bibitem[\protect\citeauthoryear{Forrey}{Forrey}{2013}]{Forrey2013}
Forrey R.~C.,  2013, \mn@doi [ApJ] {10.1088/2041-8205/773/2/L25}, 773, L25

\bibitem[\protect\citeauthoryear{{Frebel}, {Simon}, {Geha}  \&
  {Willman}}{{Frebel} et~al.}{2010}]{Frebel2010}
{Frebel} A.,  {Simon} J.~D.,  {Geha} M.,   {Willman} B.,  2010, \mn@doi [\apj]
  {10.1088/0004-637X/708/1/560}, \href
  {https://ui.adsabs.harvard.edu/abs/2010ApJ...708..560F} {708, 560}

\bibitem[\protect\citeauthoryear{{Freeman}}{{Freeman}}{1970}]{Freeman1970}
{Freeman} K.~C.,  1970, \mn@doi [\apj] {10.1086/150474}, \href
  {https://ui.adsabs.harvard.edu/abs/1970ApJ...160..811F} {160, 811}

\bibitem[\protect\citeauthoryear{Gallagher et~al.,}{Gallagher
  et~al.}{2018}]{Gallagher2018}
Gallagher M.~J.,  et~al., 2018, \mn@doi [ApJ] {10.3847/1538-4357/aabad8}, 858,
  90

\bibitem[\protect\citeauthoryear{Galli \& Palla}{Galli \&
  Palla}{1998}]{Galli.Palla1998}
Galli D.,  Palla F.,  1998, A\&A, 335, 403

\bibitem[\protect\citeauthoryear{Gao \& Solomon}{Gao \&
  Solomon}{2004}]{Gao.Solomon2004}
Gao Y.,  Solomon P.~M.,  2004, \mn@doi [ApJ] {10.1086/382999}, 606, 271

\bibitem[\protect\citeauthoryear{Genzel \& Stutzki}{Genzel \&
  Stutzki}{1989}]{Genzel.Stutzki1989}
Genzel R.,  Stutzki J.,  1989, \mn@doi [araa]
  {10.1146/annurev.aa.27.090189.000353}, 27, 41

\bibitem[\protect\citeauthoryear{{Gillmon}, {Shull}, {Tumlinson}  \&
  {Danforth}}{{Gillmon} et~al.}{2006}]{Gillmon2006}
{Gillmon} K.,  {Shull} J.~M.,  {Tumlinson} J.,   {Danforth} C.,  2006, \mn@doi
  [\apj] {10.1086/498053}, \href
  {https://ui.adsabs.harvard.edu/abs/2006ApJ...636..891G} {636, 891}

\bibitem[\protect\citeauthoryear{Glover}{Glover}{2015}]{Glover2015}
Glover S. C.~O.,  2015, \mn@doi [MNRAS] {10.1093/mnras/stv1781}, 453, 2902

\bibitem[\protect\citeauthoryear{Glover \& Abel}{Glover \&
  Abel}{2008}]{Glover.Abel2008}
Glover S. C.~O.,  Abel T.,  2008, \mn@doi [MNRAS]
  {10.1111/j.1365-2966.2008.13224.x}, 388, 1627

\bibitem[\protect\citeauthoryear{Glover \& Clark}{Glover \&
  Clark}{2012}]{Glover.Clark2012}
Glover S. C.~O.,  Clark P.~C.,  2012, \mn@doi [MNRAS]
  {10.1111/j.1365-2966.2011.20260.x}, pp no--no

\bibitem[\protect\citeauthoryear{Glover \& Jappsen}{Glover \&
  Jappsen}{2007}]{Glover.Jappsen2007}
Glover S. C.~O.,  Jappsen A.-K.,  2007, \mn@doi [ApJ] {10.1086/519445}, 666, 1

\bibitem[\protect\citeauthoryear{Glover \& Savin}{Glover \&
  Savin}{2009}]{Glover.Savin2009}
Glover S. C.~O.,  Savin D.~W.,  2009, \mn@doi [MNRAS]
  {10.1111/j.1365-2966.2008.14156.x}, 393, 911

\bibitem[\protect\citeauthoryear{Gnedin \& Kravtsov}{Gnedin \&
  Kravtsov}{2011}]{Gnedin.Kravtsov2011}
Gnedin N.~Y.,  Kravtsov A.~V.,  2011, \mn@doi [ApJ]
  {10.1088/0004-637X/728/2/88}, 728, 88

\bibitem[\protect\citeauthoryear{Gnedin, Tassis  \& Kravtsov}{Gnedin
  et~al.}{2009}]{Gnedin2009}
Gnedin N.~Y.,  Tassis K.,   Kravtsov A.~V.,  2009, \mn@doi [ApJ]
  {10.1088/0004-637X/697/1/55}, 697, 55

\bibitem[\protect\citeauthoryear{Goldsmith et~al.,}{Goldsmith
  et~al.}{2008}]{Goldsmith2008}
Goldsmith P.~F.,  et~al., 2008, \mn@doi [ApJ] {10.1086/587166}, 680, 428

\bibitem[\protect\citeauthoryear{Grassi et~al.,}{Grassi
  et~al.}{2014}]{Grassi2014}
Grassi T.,  et~al., 2014, \mn@doi [MNRAS] {10.1093/mnras/stu114}, 439, 2386

\bibitem[\protect\citeauthoryear{Grassi, Bovino, Haugboelle  \&
  Schleicher}{Grassi et~al.}{2017}]{Grassi2017}
Grassi T.,  Bovino S.,  Haugboelle T.,   Schleicher D. R.~G.,  2017, \mn@doi
  [MNRAS] {10.1093/mnras/stw2871}, 466, 1259

\bibitem[\protect\citeauthoryear{Haardt \& Madau}{Haardt \&
  Madau}{2012}]{Haardt.Madau2012}
Haardt F.,  Madau P.,  2012, \mn@doi [ApJ] {10.1088/0004-637X/746/2/125}, 746,
  125

\bibitem[\protect\citeauthoryear{Herbst}{Herbst}{2001}]{Herbst2001}
Herbst E.,  2001, \mn@doi [Chem. Soc. Rev.] {10.1039/a909040a}, 30, 168

\bibitem[\protect\citeauthoryear{Hindmarsh}{Hindmarsh}{1983}]{Hindmarsh1983}
Hindmarsh A.~C.,  1983, Scientific Computing, 1, 55

\bibitem[\protect\citeauthoryear{Hollenbach \& McKee}{Hollenbach \&
  McKee}{1979}]{Hollenbach.McKee1979}
Hollenbach D.,  McKee C.~F.,  1979, \mn@doi [ApJS] {10.1086/190631}, 41, 555

\bibitem[\protect\citeauthoryear{Hopkins}{Hopkins}{2015}]{Hopkins2015}
Hopkins P.~F.,  2015, \mn@doi [MNRAS] {10.1093/mnras/stv195}, 450, 53

\bibitem[\protect\citeauthoryear{Hopkins, Narayanan  \& Murray}{Hopkins
  et~al.}{2013}]{Hopkins2013b}
Hopkins P.~F.,  Narayanan D.,   Murray N.,  2013, \mn@doi [MNRAS]
  {10.1093/mnras/stt723}, 432, 2647

\bibitem[\protect\citeauthoryear{Hopkins et~al.,}{Hopkins
  et~al.}{2014}]{Hopkins2014}
Hopkins P.~F.,  et~al., 2014, \mn@doi [MNRAS] {10.1093/mnras/stu1738}, 445, 581

\bibitem[\protect\citeauthoryear{Hopkins et~al.,}{Hopkins
  et~al.}{2018}]{Hopkins2018}
Hopkins P.~F.,  et~al., 2018, \mn@doi [MNRAS] {10.1093/mnras/sty1690}, 480, 800

\bibitem[\protect\citeauthoryear{Hu, Naab, Walch, Moster  \& Oser}{Hu
  et~al.}{2014}]{Hu2014}
Hu C.-Y.,  Naab T.,  Walch S.,  Moster B.~P.,   Oser L.,  2014, \mn@doi [MNRAS]
  {10.1093/mnras/stu1187}, 443, 1173

\bibitem[\protect\citeauthoryear{Hu, Naab, Walch, Glover  \& Clark}{Hu
  et~al.}{2016}]{Hu2016}
Hu C.-Y.,  Naab T.,  Walch S.,  Glover S. C.~O.,   Clark P.~C.,  2016, \mn@doi
  [MNRAS] {10.1093/mnras/stw544}, 458, 3528

\bibitem[\protect\citeauthoryear{Hu, Naab, Glover, Walch  \& Clark}{Hu
  et~al.}{2017}]{Hu2017}
Hu C.-Y.,  Naab T.,  Glover S. C.~O.,  Walch S.,   Clark P.~C.,  2017, \mn@doi
  [MNRAS] {10.1093/mnras/stx1773}, 471, 2151

\bibitem[\protect\citeauthoryear{Iwamoto et~al.,}{Iwamoto
  et~al.}{1999}]{Iwamoto1999}
Iwamoto K.,  et~al., 1999, \mn@doi [ApJS] {10.1086/313278}, 125, 439

\bibitem[\protect\citeauthoryear{Jim{\'e}nez, Tissera  \&
  Matteucci}{Jim{\'e}nez et~al.}{2014}]{Jimenez2014}
Jim{\'e}nez N.,  Tissera P.~B.,   Matteucci F.,  2014, Memorie della Societa
  Astronomica Italiana, 85, 325

\bibitem[\protect\citeauthoryear{Jim{\'e}nez, Tissera  \&
  Matteucci}{Jim{\'e}nez et~al.}{2015}]{Jimenez2015}
Jim{\'e}nez N.,  Tissera P.~B.,   Matteucci F.,  2015, \mn@doi [ApJ]
  {10.1088/0004-637X/810/2/137}, 810, 137

\bibitem[\protect\citeauthoryear{Jura}{Jura}{1975}]{Jura1975}
Jura M.,  1975, \mn@doi [ApJ] {10.1086/153545}, 197, 575

\bibitem[\protect\citeauthoryear{Katz, Kimm, Sijacki  \& Haehnelt}{Katz
  et~al.}{2017}]{Katz2017}
Katz H.,  Kimm T.,  Sijacki D.,   Haehnelt M.,  2017, \mn@doi [MNRAS]
  {10.1093/mnras/stx608}, 468, 4831

\bibitem[\protect\citeauthoryear{Kennicutt}{Kennicutt}{1989}]{Kennicutt1989}
Kennicutt Jr. R.~C.,  1989, \mn@doi [ApJ] {10.1086/167834}, 344, 685

\bibitem[\protect\citeauthoryear{Kennicutt}{Kennicutt}{1998}]{Kennicutt1998}
Kennicutt Jr. R.~C.,  1998, ApJ, 498, 541

\bibitem[\protect\citeauthoryear{Kennicutt Jr. et~al.,}{Kennicutt
  et~al.}{2007}]{Kennicutt2007}
Kennicutt Jr. R.~C.,  et~al., 2007, \mn@doi [ApJ] {10.1086/522300}, 671, 333

\bibitem[\protect\citeauthoryear{Krumholz \& Gnedin}{Krumholz \&
  Gnedin}{2011}]{Krumholz.Gnedin2011}
Krumholz M.~R.,  Gnedin N.~Y.,  2011, \mn@doi [ApJ]
  {10.1088/0004-637X/729/1/36}, 729, 36

\bibitem[\protect\citeauthoryear{Krumholz \& McKee}{Krumholz \&
  McKee}{2005}]{Krumholz.McKee2005}
Krumholz M.~R.,  McKee C.~F.,  2005, \mn@doi [ApJ] {10.1086/431734}, 630, 250

\bibitem[\protect\citeauthoryear{Krumholz, McKee  \& Tumlinson}{Krumholz
  et~al.}{2009}]{Krumholz2009}
Krumholz M.~R.,  McKee C.~F.,   Tumlinson J.,  2009, \mn@doi [ApJ]
  {10.1088/0004-637X/699/1/850}, 699, 850

\bibitem[\protect\citeauthoryear{Krumholz, Dekel  \& McKee}{Krumholz
  et~al.}{2012}]{Krumholz2012}
Krumholz M.~R.,  Dekel A.,   McKee C.~F.,  2012, \mn@doi [ApJ]
  {10.1088/0004-637X/745/1/69}, 745, 69

\bibitem[\protect\citeauthoryear{Kuhlen, Krumholz, Madau, Smith  \&
  Wise}{Kuhlen et~al.}{2012}]{Kuhlen2012}
Kuhlen M.,  Krumholz M.~R.,  Madau P.,  Smith B.~D.,   Wise J.,  2012, \mn@doi
  [ApJ] {10.1088/0004-637X/749/1/36}, 749, 36

\bibitem[\protect\citeauthoryear{Larson}{Larson}{1981}]{Larson1981}
Larson R.~B.,  1981, \mn@doi [MNRAS] {10.1093/mnras/194.4.809}, 194, 809

\bibitem[\protect\citeauthoryear{{Latif} \& {Schleicher}}{{Latif} \&
  {Schleicher}}{2015}]{Latif.Schleicher2015}
{Latif} M.~A.,  {Schleicher} D.~R.~G.,  2015, \mn@doi [\mnras]
  {10.1093/mnras/stu2573}, \href
  {https://ui.adsabs.harvard.edu/abs/2015MNRAS.449...77L} {449, 77}

\bibitem[\protect\citeauthoryear{{Latif} et~al.,}{{Latif}
  et~al.}{2019}]{Latif2019}
{Latif} M.~A.,  et~al., 2019, \mn@doi [\mnras] {10.1093/mnras/stz608}, \href
  {https://ui.adsabs.harvard.edu/abs/2019MNRAS.485.3352L} {485, 3352}

\bibitem[\protect\citeauthoryear{Leroy et~al.,}{Leroy et~al.}{2008}]{Leroy2008}
Leroy A.~K.,  et~al., 2008, \mn@doi [AJ] {10.1088/0004-6256/136/6/2782}, 136,
  2782

\bibitem[\protect\citeauthoryear{Leroy et~al.,}{Leroy et~al.}{2013}]{Leroy2013}
Leroy A.~K.,  et~al., 2013, \mn@doi [AJ] {10.1088/0004-6256/146/2/19}, 146, 19

\bibitem[\protect\citeauthoryear{{Lupi} \& {Bovino}}{{Lupi} \&
  {Bovino}}{2020}]{Lupi.Bovino2020}
{Lupi} A.,  {Bovino} S.,  2020, \mn@doi [\mnras] {10.1093/mnras/staa048}, \href
  {https://ui.adsabs.harvard.edu/abs/2020MNRAS.492.2818L} {492, 2818}

\bibitem[\protect\citeauthoryear{Lupi, Bovino, Capelo, Volonteri  \& Silk}{Lupi
  et~al.}{2018}]{Lupi2018}
Lupi A.,  Bovino S.,  Capelo P.~R.,  Volonteri M.,   Silk J.,  2018, \mn@doi
  [MNRAS] {10.1093/mnras/stx2874}, 474, 2884

\bibitem[\protect\citeauthoryear{{Mac Low} \& {Glover}}{{Mac Low} \&
  {Glover}}{2012}]{MacLow.Glover2012}
{Mac Low} M.-M.,  {Glover} S. C.~O.,  2012, \mn@doi [\apj]
  {10.1088/0004-637X/746/2/135}, \href
  {https://ui.adsabs.harvard.edu/abs/2012ApJ...746..135M} {746, 135}

\bibitem[\protect\citeauthoryear{Maio, Dolag, Ciardi  \& Tornatore}{Maio
  et~al.}{2007}]{Maio2007}
Maio U.,  Dolag K.,  Ciardi B.,   Tornatore L.,  2007, \mn@doi [MNRAS]
  {10.1111/j.1365-2966.2007.12016.x}, 379, 963

\bibitem[\protect\citeauthoryear{{Maio}, {Ciardi}, {Dolag}, {Tornatore}  \&
  {Khochfar}}{{Maio} et~al.}{2010}]{Maio2010}
{Maio} U.,  {Ciardi} B.,  {Dolag} K.,  {Tornatore} L.,   {Khochfar} S.,  2010,
  \mn@doi [\mnras] {10.1111/j.1365-2966.2010.17003.x}, \href
  {https://ui.adsabs.harvard.edu/abs/2010MNRAS.407.1003M} {407, 1003}

\bibitem[\protect\citeauthoryear{Maiolino et~al.,}{Maiolino
  et~al.}{2008}]{Maiolino2008}
Maiolino R.,  et~al., 2008, \mn@doi [A\&A] {10.1051/0004-6361:200809678}, 488,
  463

\bibitem[\protect\citeauthoryear{Marri \& White}{Marri \&
  White}{2003}]{Marri.White2003}
Marri S.,  White S. D.~M.,  2003, \mn@doi [Monthly Notices RAS]
  {10.1046/j.1365-8711.2003.06984.x}, 345, 561

\bibitem[\protect\citeauthoryear{Matteucci \& Recchi}{Matteucci \&
  Recchi}{2001}]{Matteucci.Recchi2001}
Matteucci F.,  Recchi S.,  2001, \mn@doi [ApJ] {10.1086/322472}, 558, 351

\bibitem[\protect\citeauthoryear{McKee \& Krumholz}{McKee \&
  Krumholz}{2010}]{McKee.Krumholz2010}
McKee C.~F.,  Krumholz M.~R.,  2010, \mn@doi [ApJ]
  {10.1088/0004-637X/709/1/308}, 709, 308

\bibitem[\protect\citeauthoryear{McKee \& Ostriker}{McKee \&
  Ostriker}{2007}]{McKee.Ostriker2007}
McKee C.~F.,  Ostriker E.~C.,  2007, \mn@doi [Annu. Rev. Astron. Astrophys.]
  {10.1146/annurev.astro.45.051806.110602}, 45, 565

\bibitem[\protect\citeauthoryear{Micic, Glover, Federrath  \& Klessen}{Micic
  et~al.}{2012}]{Micic2012}
Micic M.,  Glover S. C.~O.,  Federrath C.,   Klessen R.~S.,  2012, \mn@doi
  [MNRAS] {10.1111/j.1365-2966.2012.20477.x}, 421, 2531

\bibitem[\protect\citeauthoryear{Mosconi, Tissera, Lambas  \& Cora}{Mosconi
  et~al.}{2001}]{Mosconi2001}
Mosconi M.~B.,  Tissera P.~B.,  Lambas D.~G.,   Cora S.~A.,  2001, \mn@doi
  [MNRAS] {10.1046/j.1365-8711.2001.04198.x}, 325, 34

\bibitem[\protect\citeauthoryear{Navarro, Frenk  \& White}{Navarro
  et~al.}{1996}]{NFW1996}
Navarro J.~F.,  Frenk C.~S.,   White S. D.~M.,  1996, \mn@doi [ApJ]
  {10.1086/177173}, 462, 563

\bibitem[\protect\citeauthoryear{Nickerson, Teyssier  \& Rosdahl}{Nickerson
  et~al.}{2018}]{Nickerson2018}
Nickerson S.,  Teyssier R.,   Rosdahl J.,  2018, \mn@doi [MNRAS]
  {10.1093/mnras/sty1556}, 479, 3206

\bibitem[\protect\citeauthoryear{Nickerson, Teyssier  \& Rosdahl}{Nickerson
  et~al.}{2019}]{Nickerson2019}
Nickerson S.,  Teyssier R.,   Rosdahl J.,  2019, \mn@doi [MNRAS]
  {10.1093/mnras/stz048}

\bibitem[\protect\citeauthoryear{Omukai}{Omukai}{2000}]{Omukai2000}
Omukai K.,  2000, ApJ, 534, 809

\bibitem[\protect\citeauthoryear{Pallottini et~al.,}{Pallottini
  et~al.}{2017}]{Pallottini2017}
Pallottini A.,  et~al., 2017, \mn@doi [MNRAS] {10.1093/mnras/stx1792}, 471,
  4128

\bibitem[\protect\citeauthoryear{Pilbratt et~al.,}{Pilbratt
  et~al.}{2010}]{Pilbratt2010}
Pilbratt G.~L.,  et~al., 2010, \mn@doi [A\&A] {10.1051/0004-6361/201014759},
  518, L1

\bibitem[\protect\citeauthoryear{Pineda et~al.,}{Pineda
  et~al.}{2010}]{Pineda2010}
Pineda J.~L.,  et~al., 2010, \mn@doi [ApJ] {10.1088/0004-637X/721/1/686}, 721,
  686

\bibitem[\protect\citeauthoryear{{Plat} et~al.,}{{Plat}
  et~al.}{2019}]{Plat2019}
{Plat} A.,  et~al., 2019, \mn@doi [\mnras] {10.1093/mnras/stz2616}, \href
  {https://ui.adsabs.harvard.edu/abs/2019MNRAS.490..978P} {490, 978}

\bibitem[\protect\citeauthoryear{Price}{Price}{2008}]{Price2008}
Price D.~J.,  2008, \mn@doi [J. Comput. Phys.] {10.1016/j.jcp.2008.08.011},
  227, 10040

\bibitem[\protect\citeauthoryear{Richings \& Schaye}{Richings \&
  Schaye}{2016a}]{Richings.Schaye2016a}
Richings A.~J.,  Schaye J.,  2016a, \mn@doi [MNRAS] {10.1093/mnras/stw327},
  458, 270

\bibitem[\protect\citeauthoryear{Richings \& Schaye}{Richings \&
  Schaye}{2016b}]{Richings.Schaye2016b}
Richings A.~J.,  Schaye J.,  2016b, \mn@doi [MNRAS] {10.1093/mnras/stw1135},
  460, 2297

\bibitem[\protect\citeauthoryear{Richings, Schaye  \& Oppenheimer}{Richings
  et~al.}{2014a}]{Richings2014a}
Richings A.~J.,  Schaye J.,   Oppenheimer B.~D.,  2014a, \mn@doi [MNRAS]
  {10.1093/mnras/stu525}, 440, 3349

\bibitem[\protect\citeauthoryear{Richings, Schaye  \& Oppenheimer}{Richings
  et~al.}{2014b}]{Richings2014b}
Richings A.~J.,  Schaye J.,   Oppenheimer B.~D.,  2014b, \mn@doi [MNRAS]
  {10.1093/mnras/stu1046}, 442, 2780

\bibitem[\protect\citeauthoryear{Robertson \& Kravtsov}{Robertson \&
  Kravtsov}{2008}]{Robertson.Kravtsov2008}
Robertson B.~E.,  Kravtsov A.~V.,  2008, \mn@doi [ApJ] {10.1086/587796}, 680,
  1083

\bibitem[\protect\citeauthoryear{{Safranek-Shrader} et~al.,}{{Safranek-Shrader}
  et~al.}{2017}]{Safranek-Shrader2017}
{Safranek-Shrader} C.,  et~al., 2017, \mn@doi [MNRAS] {10.1093/mnras/stw2647},
  465, 885

\bibitem[\protect\citeauthoryear{Saintonge et~al.,}{Saintonge
  et~al.}{2017}]{Saintonge2017}
Saintonge A.,  et~al., 2017, \mn@doi [ApJS] {10.3847/1538-4365/aa97e0}, 233, 22

\bibitem[\protect\citeauthoryear{{Salpeter}}{{Salpeter}}{1955}]{Salpeter1955}
{Salpeter} E.~E.,  1955, \mn@doi [\apj] {10.1086/145971}, \href
  {https://ui.adsabs.harvard.edu/abs/1955ApJ...121..161S} {121, 161}

\bibitem[\protect\citeauthoryear{{S{\'a}nchez-Bl{\'a}zquez}
  et~al.,}{{S{\'a}nchez-Bl{\'a}zquez} et~al.}{2014}]{Sanchez-Blazquez2014}
{S{\'a}nchez-Bl{\'a}zquez} P.,  et~al., 2014, \mn@doi [A\&A]
  {10.1051/0004-6361/201423635}, 570, A6

\bibitem[\protect\citeauthoryear{Scannapieco, Tissera, White  \&
  Springel}{Scannapieco et~al.}{2005}]{Scannapieco2005}
Scannapieco C.,  Tissera P.~B.,  White S. D.~M.,   Springel V.,  2005, \mn@doi
  [MNRAS] {10.1111/j.1365-2966.2005.09574.x}, 364, 552

\bibitem[\protect\citeauthoryear{Scannapieco, Tissera, White  \&
  Springel}{Scannapieco et~al.}{2006}]{Scannapieco2006}
Scannapieco C.,  Tissera P.~B.,  White S. D.~M.,   Springel V.,  2006, \mn@doi
  [MNRAS] {10.1111/j.1365-2966.2006.10785.x}, 371, 1125

\bibitem[\protect\citeauthoryear{Scannapieco, Tissera, White  \&
  Springel}{Scannapieco et~al.}{2008}]{Scannapieco2008}
Scannapieco C.,  Tissera P.~B.,  White S. D.~M.,   Springel V.,  2008, \mn@doi
  [MNRAS] {10.1111/j.1365-2966.2008.13678.x}, 389, 1137

\bibitem[\protect\citeauthoryear{Schmidt}{Schmidt}{1963}]{Schmidt1963}
Schmidt M.,  1963, \mn@doi [ApJ] {10.1086/147553}, 137, 758

\bibitem[\protect\citeauthoryear{Schmidt, Wilson  \& Observatories}{Schmidt
  et~al.}{1959}]{Schmidt1959}
Schmidt M.,  Wilson M.,   Observatories P.,  1959, ApJ, 129, 16

\bibitem[\protect\citeauthoryear{{Schneider}, {Omukai}, {Inoue}  \&
  {Ferrara}}{{Schneider} et~al.}{2006}]{Schneider2006}
{Schneider} R.,  {Omukai} K.,  {Inoue} A.~K.,   {Ferrara} A.,  2006, \mn@doi
  [\mnras] {10.1111/j.1365-2966.2006.10391.x}, \href
  {https://ui.adsabs.harvard.edu/abs/2006MNRAS.369.1437S} {369, 1437}

\bibitem[\protect\citeauthoryear{Schruba et~al.,}{Schruba
  et~al.}{2011}]{Schruba2011}
Schruba A.,  et~al., 2011, \mn@doi [AJ] {10.1088/0004-6256/142/2/37}, 142, 37

\bibitem[\protect\citeauthoryear{Schuster, Kramer, Hitschfeld, {Garcia-Burillo}
   \& Mookerjea}{Schuster et~al.}{2007}]{Schuster2007}
Schuster K.~F.,  Kramer C.,  Hitschfeld M.,  {Garcia-Burillo} S.,   Mookerjea
  B.,  2007, \mn@doi [A\&A] {10.1051/0004-6361:20065579}, 461, 143

\bibitem[\protect\citeauthoryear{{Seifried}, {S{\'a}nchez-Monge}, {Suri}  \&
  {Walch}}{{Seifried} et~al.}{2017}]{Seifried2017}
{Seifried} D.,  {S{\'a}nchez-Monge} {\'A}.,  {Suri} S.,   {Walch} S.,  2017,
  \mn@doi [\mnras] {10.1093/mnras/stx399}, \href
  {https://ui.adsabs.harvard.edu/abs/2017MNRAS.467.4467S} {467, 4467}

\bibitem[\protect\citeauthoryear{Semenov, Kravtsov  \& Gnedin}{Semenov
  et~al.}{2016}]{Semenov2016}
Semenov V.~A.,  Kravtsov A.~V.,   Gnedin N.~Y.,  2016, \mn@doi [ApJ]
  {10.3847/0004-637X/826/2/200}, 826, 200

\bibitem[\protect\citeauthoryear{{Sharda}, {Krumholz}  \& {Federrath}}{{Sharda}
  et~al.}{2019}]{Piyush2019}
{Sharda} P.,  {Krumholz} M.~R.,   {Federrath} C.,  2019, \mn@doi [\mnras]
  {10.1093/mnras/stz2618}, \href
  {https://ui.adsabs.harvard.edu/abs/2019MNRAS.490..513S} {490, 513}

\bibitem[\protect\citeauthoryear{Shen, Wadsley  \& Stinson}{Shen
  et~al.}{2010}]{Shen2010}
Shen S.,  Wadsley J.,   Stinson G.,  2010, \mn@doi [MNRAS]
  {10.1111/j.1365-2966.2010.17047.x}, 407, 1581

\bibitem[\protect\citeauthoryear{Shen et~al.,}{Shen et~al.}{2013}]{Shen2013}
Shen S.,  et~al., 2013, \mn@doi [ApJ] {10.1088/0004-637X/765/2/89}, 765, 89

\bibitem[\protect\citeauthoryear{{Shull}, {Danforth}  \& {Anderson}}{{Shull}
  et~al.}{2021}]{Shull2021}
{Shull} J.~M.,  {Danforth} C.~W.,   {Anderson} K.~L.,  2021, arXiv e-prints,
  \href {https://ui.adsabs.harvard.edu/abs/2021arXiv210211301S} {p.
  arXiv:2102.11301}

\bibitem[\protect\citeauthoryear{Solomon, Rivolo, Barrett  \& Yahil}{Solomon
  et~al.}{1987}]{Solomon1987}
Solomon P.~M.,  Rivolo A.~R.,  Barrett J.,   Yahil A.,  1987, \mn@doi [ApJ]
  {10.1086/165493}, 319, 730

\bibitem[\protect\citeauthoryear{Springel}{Springel}{2005}]{Springel2005b}
Springel V.,  2005, \mn@doi [MNRAS] {10.1111/j.1365-2966.2005.09655.x}, 364,
  1105

\bibitem[\protect\citeauthoryear{Springel \& Hernquist}{Springel \&
  Hernquist}{2003}]{Springel.Hernquist2003}
Springel V.,  Hernquist L.,  2003, \mn@doi [MNRAS]
  {10.1046/j.1365-8711.2003.06206.x}, 339, 289

\bibitem[\protect\citeauthoryear{Stecher \& Williams}{Stecher \&
  Williams}{1967}]{Stecher.Williams1967}
Stecher T.~P.,  Williams D.~A.,  1967, \mn@doi [ApJ] {10.1086/180047}, 149, L29

\bibitem[\protect\citeauthoryear{Sternberg, Le~Petit, Roueff  \&
  Le~Bourlot}{Sternberg et~al.}{2014}]{Sternberg2014}
Sternberg A.,  Le~Petit F.,  Roueff E.,   Le~Bourlot J.,  2014, \mn@doi [ApJ]
  {10.1088/0004-637X/790/1/10}, 790, 10

\bibitem[\protect\citeauthoryear{Strong \& Mattox}{Strong \&
  Mattox}{1996}]{Strong.Mattox1996}
Strong A.~W.,  Mattox J.~R.,  1996, A\&A, 308, 21

\bibitem[\protect\citeauthoryear{{Tielens} \& {Hollenbach}}{{Tielens} \&
  {Hollenbach}}{1985}]{Tielens.Hollenbach1985}
{Tielens} A.~G.~G.~M.,  {Hollenbach} D.,  1985, \mn@doi [\apj]
  {10.1086/163111}, \href
  {https://ui.adsabs.harvard.edu/abs/1985ApJ...291..722T} {291, 722}

\bibitem[\protect\citeauthoryear{Tissera, Pedrosa, Sillero  \& Vilchez}{Tissera
  et~al.}{2016a}]{Tissera2016a}
Tissera P.~B.,  Pedrosa S.~E.,  Sillero E.,   Vilchez J.~M.,  2016a, \mn@doi
  [MNRAS] {10.1093/mnras/stv2736}, 456, 2982

\bibitem[\protect\citeauthoryear{Tissera et~al.,}{Tissera
  et~al.}{2016b}]{Tissera2016}
Tissera P.~B.,  et~al., 2016b, \mn@doi [A\&A] {10.1051/0004-6361/201628188},
  592, A93

\bibitem[\protect\citeauthoryear{Tomassetti, Porciani, {Romano-Diaz}  \&
  Ludlow}{Tomassetti et~al.}{2014}]{Tomassetti2014}
Tomassetti M.,  Porciani C.,  {Romano-Diaz} E.,   Ludlow A.~D.,  2014, \mn@doi
  [MNRAS] {10.1093/mnras/stu2273}, 446, 3330

\bibitem[\protect\citeauthoryear{Tricco \& Price}{Tricco \&
  Price}{2013}]{Tricco.Price2013}
Tricco T.~S.,  Price D.~J.,  2013, \mn@doi [MNRAS] {10.1093/mnras/stt1776},
  436, 2810

\bibitem[\protect\citeauthoryear{Tumlinson}{Tumlinson}{2002}]{Tumlinson2002}
Tumlinson J.,  2002, \mn@doi [ApJ] {10.1086/338112}, 566, 857

\bibitem[\protect\citeauthoryear{Wakelam et~al.,}{Wakelam
  et~al.}{2017}]{Wakelam2017}
Wakelam V.,  et~al., 2017, arXiv:1711.10568 [astro-ph]

\bibitem[\protect\citeauthoryear{{Weingartner} \& {Draine}}{{Weingartner} \&
  {Draine}}{2001}]{Weingartner.Draine2001}
{Weingartner} J.~C.,  {Draine} B.~T.,  2001, \mn@doi [\apj] {10.1086/318651},
  \href {https://ui.adsabs.harvard.edu/abs/2001ApJ...548..296W} {548, 296}

\bibitem[\protect\citeauthoryear{Williams \& McKee}{Williams \&
  McKee}{1997}]{Williams.McKee1997}
Williams J.~P.,  McKee C.~F.,  1997, \mn@doi [ApJ] {10.1086/303588}, 476, 166

\bibitem[\protect\citeauthoryear{Wolfire, McKee, Hollenbach  \&
  Tielens}{Wolfire et~al.}{2003}]{Wolfire2003}
Wolfire M.~G.,  McKee C.~F.,  Hollenbach D.,   Tielens A. G. G.~M.,  2003,
  \mn@doi [ApJ] {10.1086/368016}, 587, 278

\bibitem[\protect\citeauthoryear{Wolfire, Tielens, Hollenbach  \&
  Kaufman}{Wolfire et~al.}{2008}]{Wolfire2008}
Wolfire M.~G.,  Tielens A. G. G.~M.,  Hollenbach D.,   Kaufman M.~J.,  2008,
  \mn@doi [ApJ] {10.1086/587688}, 680, 384

\bibitem[\protect\citeauthoryear{Wong \& Blitz}{Wong \&
  Blitz}{2002}]{Wong.Blitz2002}
Wong T.,  Blitz L.,  2002, \mn@doi [ApJ] {10.1086/339287}, 569, 157

\bibitem[\protect\citeauthoryear{Woosley \& Weaver}{Woosley \&
  Weaver}{1995}]{Woosley.Weaver1995}
Woosley S.~E.,  Weaver T.~A.,  1995, ApJ Supplement Series, p.~55

\bibitem[\protect\citeauthoryear{Yamasawa et~al.,}{Yamasawa
  et~al.}{2011}]{Yamasawa2011}
Yamasawa D.,  et~al., 2011, \mn@doi [ApJ] {10.1088/0004-637X/735/1/44}, 735, 44

\makeatother
\end{thebibliography}


\bsp	
\label{lastpage}
\end{document}



\section*{Supplementary online material}
\vspace{1cm}
\begin{figure*}[h]
    \centering
	\includegraphics[trim={0 25 0 0}, width=\textwidth]{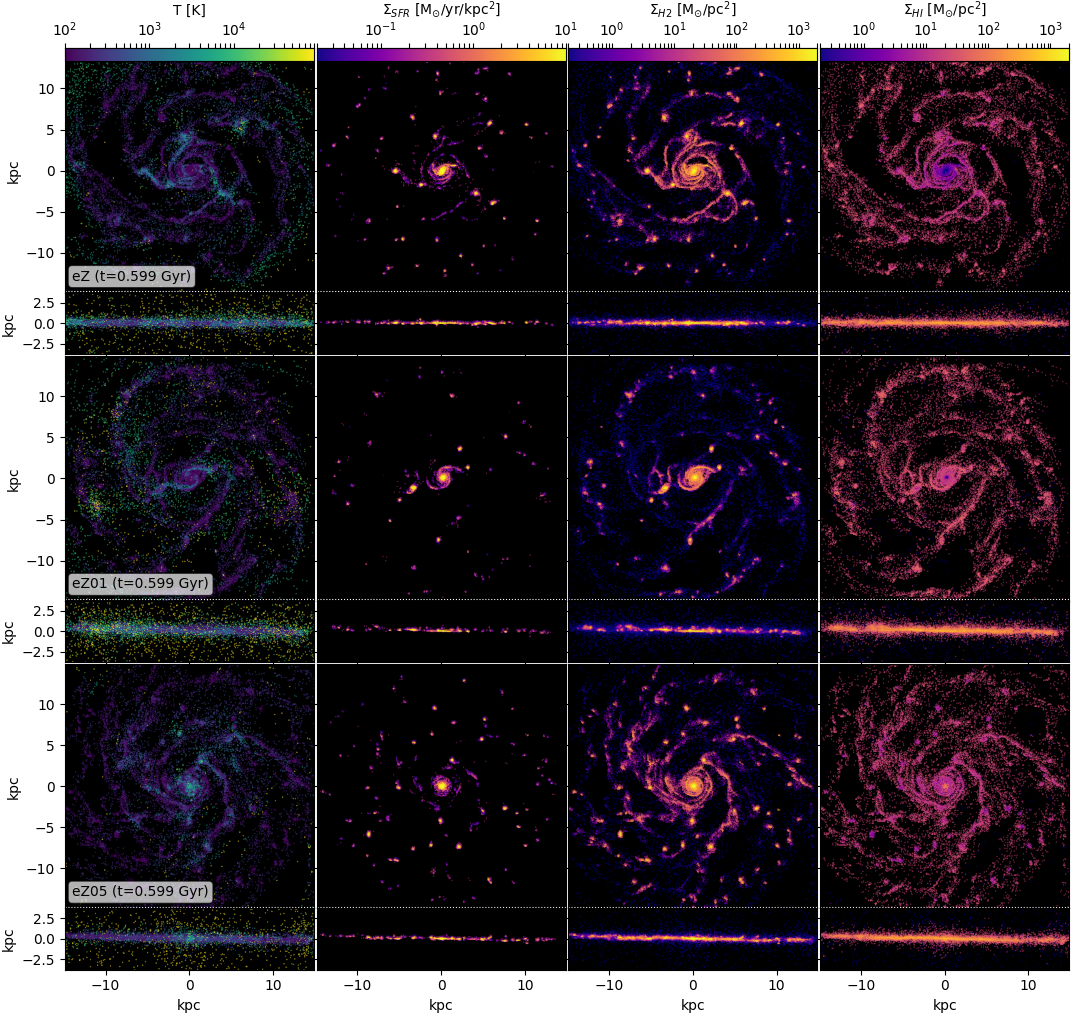}
    \vspace{0.2cm}
    \caption{Face-on and edge-on maps (small panels) of the mass-weighted gas temperature (first column) and the projected surface density of SFR (second column), H$_2$ (third column) and HI (fourth) content in eZ (fist row), eZ01 (second row) and eZ05 (third row).}
    \label{fig:d_colormap_1}
\end{figure*}